\documentclass[journal,comsoc]{IEEEtran}
\usepackage[T1]{fontenc}% optional T1 font encoding

\ifCLASSINFOpdf
   \usepackage[pdftex]{graphicx}
\else
   or other class option (dvipsone, dvipdf, if not using dvips). graphicx
   \usepackage[dvips]{graphicx}
   \graphicspath{{../eps/}}
   \DeclareGraphicsExtensions{.eps}
\fi

\usepackage{tabularx}
\usepackage{multirow}
\usepackage{booktabs}
\usepackage{array}
\newcolumntype{L}[1]{>{\raggedright\let\newline\\\arraybackslash\hspace{0pt}}m{#1}}
\newcolumntype{C}[1]{>{\centering\let\newline\\\arraybackslash\hspace{0pt}}m{#1}}
\newcolumntype{R}[1]{>{\raggedleft\let\newline\\\arraybackslash\hspace{0pt}}m{#1}}
\usepackage{tikz}

\usepackage[english]{babel}
\usepackage{amsthm}
\usepackage{amssymb}

\newtheorem{theorem}{Theorem}
\newtheorem{lemma}{Lemma}

\theoremstyle{plain}
\newtheorem{proposition}{Proposition}
\theoremstyle{remark}

\usepackage{cite}

\usepackage[normalem]{ulem}

\usepackage{amsmath}
\usepackage[cmintegrals]{newtxmath}
\usepackage{wasysym}

\usepackage{algorithmic}
\usepackage{algorithm}
\usepackage{comment}

\usepackage{verbatim}

\usepackage{array}
\usepackage{makecell}
\usepackage[pagewise]{lineno}
\usepackage{upgreek}

\ifCLASSOPTIONcompsoc
  \usepackage[caption=false,font=normalsize,uabelfont=sf,textfont=sf]{subfig}
\else
  \usepackage[caption=false,font=footnotesize]{subfig}
\fi
\begin{document}
% \linenumbers
\setcounter{figure}{0}
\renewcommand{\figurename}{Fig.}
\renewcommand{\thefigure}{\arabic{figure}}
% paper title
% Titles are generally capitalized except for words such as a, an, and, as,
% at, but, by, for, in, nor, of, on, or, the, to and up, which are usually
% not capitalized unless they are the first or last word of the title.
% Linebreaks \\ can be used within to get better formatting as desired.
% Do not put math or special symbols in the title.
\title{\textcolor{black}{Signaling Design for Noncoherent Distributed Integrated Sensing and Communication Systems}}
%\pagestyle{empty}
%
% author names and IEEE memberships
% note positions of commas and nonbreaking spaces ( ~ ) LaTeX will not break
% a structure at a ~ so this keeps an author's name from being broken across
% two lines.
% use \thanks{} to gain access to the first footnote area
% a separate \thanks must be used for each paragraph as LaTeX2e's \thanks
% was not built to handle multiple paragraphs
%

\author{Kawon~Han,~\IEEEmembership{Member,~IEEE,} Kaitao~Meng,~\IEEEmembership{Member,~IEEE,} and Christos~Masouros,~\IEEEmembership{Fellow,~IEEE}

\thanks{Manuscript received xx, 2024. K. Han, K. Meng, and C. Masouros are with the Department of Electronic and Electrical Engineering, University College London, London, UK (emails: {kawon.han, kaitao.meng, c.masouros}@ucl.ac.uk).}}

\maketitle

% As a general rule, do not put math, special symbols or citations
% in the abstract or keywords.
\begin{abstract}
\textcolor{black}{The ultimate goal of enabling sensing through the cellular network is to obtain coordinated sensing of an unprecedented scale, through distributed integrated sensing and communication (D-ISAC). This, however, introduces challenges related to synchronization and demands new transmission methodologies.} In this paper, we propose a transmit signal design framework for D-ISAC systems, where multiple ISAC nodes cooperatively perform sensing and communication without requiring phase-level synchronization. The proposed framework \textcolor{black}{employing orthogonal frequency division multiplexing (OFDM)} jointly designs downlink coordinated multi-point (CoMP) communication signals and multi-input multi-output (MIMO) radar signals, leveraging both collocated and distributed MIMO radars to estimate angle-of-arrival (AOA) and time-of-flight (TOF) from all possible multi-static measurements for target localization. To design the optimal D-ISAC transmit signal, we use the target localization Cramér-Rao bound (CRB) as the sensing performance metric and the signal-to-interference-plus-noise ratio (SINR) as the communication performance metric. \textcolor{black}{Then, an optimization problem} is formulated to minimize the localization CRB while maintaining a minimum SINR requirement for each communication user. Moreover, we present three distinct transmit signal design approaches, including optimal, orthogonal, and beamforming designs, which reveal trade-offs between ISAC performance and computational complexity. Unlike single-node ISAC systems, the proposed D-ISAC designs involve per-subcarrier sensing signal optimization to enable accurate TOF estimation, which contributes to the target localization performance. \textcolor{black}{Numerical simulations demonstrate the effectiveness of the proposed designs in achieving flexible ISAC trade-offs and efficient D-ISAC signal transmission.}
\end{abstract}


% For peer review papers, you can put extra information on the cover
% page as needed:
% \ifCLASSOPTIONpeerreview
% \begin{center} \bfseries EDICS Category: 3-BBND \end{center}
% \fi
%
% For peerreview papers, this IEEEtran command inserts a page break and
% creates the second title. It will be ignored for other modes.
\IEEEpeerreviewmaketitle

\section{Introduction}
\IEEEPARstart{W}ith the exponential growth in demand for the efficient-use of wireless resources, integrated sensing and communication (ISAC) technologies have emerged as an innovative technology, garnering significant attention across academic and industrial sectors \cite{ISAC1, ISAC2, zhang2021enabling}. ISAC systems unify radar sensing and communication functionalities within a shared framework, leveraging common hardware and resources to achieve dual-purpose operations. This integrated approach not only optimizes spectral utilization but also substantially reduces hardware costs and energy consumption, making it a highly efficient alternative to the traditional independent deployment of communication and radar systems \cite{ISAC3, ISAC4}.

Recent advancements in ISAC have predominantly focused on signal design \cite{he2023full,li2017joint,nguyen2023multiuser, liu2018toward} and resource allocation strategies \cite{valiulahi2023net, dong2022sensing} for single-node configurations. These systems typically employ monostatic radar sensing and single basestation (BS) communication services, offering the enhanced spectrum efficiency and the improved ISAC performance. \textcolor{black}{However, the ultimate goal of the ISAC deployment in cellular systems is to enable networked sensing of an unprecedented scale. A significant barrier in achieving this is the synchronization between distributed nodes. Moreover,} the increasing network density in next-generation wireless systems causes challenges such as inter-node interference and resource congestion \cite{wei2024deep, strinati2024towards}. Therefore, the exploitation of networked ISAC systems and cooperation of multiple cells has been emerging as a promising configuration of the future wireless system .

Unlike single-node systems, networked ISAC involves multiple interconnected nodes or BSs that operate in \textcolor{black}{adjacent locations}, which provide cooperative gains that significantly enhance both communication and sensing capabilities \cite{zhu2024enabling, thoma2023distributed}. By leveraging collaborative capabilities, they utilizes multiple communication and sensing links, mitigating interference, and exploiting spatial diversity. This cooperative approach facilitates enhanced spectral efficiency, \textcolor{black}{broader communication and sensing coverage}, and enhanced accuracy in target parameter estimation, unlocking a new opportunity in ISAC technology. Recent investigations into network-level cooperative ISAC systems have explored areas such as spatial resource allocation \cite{meng2024cooperative}, cooperative cluster configurations \cite{meng2024bs}, and antenna-to-BS allocation \cite{meng2024network}, demonstrating the performance benefits achievable through multi-node cooperation. 

Recently, system-level architectures for cooperative and networked ISAC systems have been studied across various scenarios to explore the potential of distributed configurations for ISAC \cite{lou2024beamforming,   jiang2024cooperation, chowdary2024hybrid, zhang2024target, wei2024integrated, yang2024coordinated, ji2023networking, wang2024collaborative}. In \cite{lou2024beamforming,  jiang2024cooperation}, the cooperation of active and passive sensing has been investigated, where sensing signals received at multiple receivers are used to localize targets, corresponding to distributed single-input multi-output (SIMO) radar. Although they leverage increased spatial diversity by distributed architectures, those systems cannot fully enjoy cooperative transmission gains for both communication and sensing. On the other hand, \cite{zhang2024target, wei2024integrated, chowdary2024hybrid} considered  sensing cooperation based on distributed multi-input single-output (MISO) systems, leveraging transmitted signals from multiple transmit nodes, which lack of considerations of receiver diversity to improve distributed sensing performance. Distributed architectures presented in \cite{yang2024coordinated, ji2023networking, wang2024collaborative} explore joint collaboration among multiple transmitters and receivers to estimate target locations while simultaneously serving communication users or mitigating inter-node interference. However, these works fall short of fully exploiting the potential gains distributed ISAC (D-ISAC) achievable through spatial diversity of distributed antenna arrays, because they exploit only parts of scattered signals as only bistatic sensing \cite{yang2024coordinated} or only monostatic sensing \cite{ji2023networking, wang2024collaborative}.

\textcolor{black}{While preliminary studies have investigated various levels of cooperation with distributed architectures, signal-level designs remain underdeveloped for cooperative ISAC systems. In terms of transmit beamforming designs, \cite{lou2024beamforming} focused solely on multi-static sensing with distributed receivers, neglecting the distributed transmission in D-ISAC. On the other hand, \cite{gao2023cooperative} investigated cooperative beamforming in distributed cloud radio access networks (C-RAN) integrated with rate-splitting multiple access. Likewise, \cite{yang2024coordinated} proposed coordinated transmit beamforming for distributed sensing and multi-user communication, emphasizing inter-cell interference mitigation. Although these coordinated beamforming approaches mitigate inter-node interference induced by simultaneous signal transmission from multiple nodes, they cannot exploit the signal and/or power combining gain in the D-ISAC system.}

\textcolor{black}{On the other hand, D-ISAC systems that utilize joint transmission schemes by sharing the transmitted signals among cooperative nodes have been considered in \cite{xu2023integrated, demirhan2024cell, babu2024precoding}, which further improves the cooperative gain compared to D-ISAC using coordinated beamforming. In \cite{xu2023integrated}, a beamforming design for multi-static sensing and joint transmission coordinated multipoint (CoMP) communication has been investigated. A related work \cite{demirhan2024cell} introduced a transmit beamforming approach for cell-free multi-input multi-output (MIMO) ISAC systems, coherently serving communication users with multiple access points. In multi-cell ISAC \cite{babu2024precoding}, both block- and symbol-level precoding designs with joint transmission CoMP have been considered. However, these systems employ the coherent cooperation with phase-level synchronization among distributed nodes, which is challenging to achieve in practice. Also, they do not take the target localization gain into account in the signaling design, which is the unique advantage of the use of distributed MIMO radar\cite{godrich2010target}.}

\textcolor{black}{Although classical coordinated beamforming \cite{dahrouj2010coordinated} and CoMP transmission \cite{irmer2011coordinated} can significantly enhance cooperative communication performance compared to the single-cell system, these approaches do not fully exploit the sensing gains achievable through distributed MIMO radar. Prior works discussed above primarily consider the improvement of AOA measurement, focusing on only beamforming design for their transmit signaling. Interestingly, distributed radar sensing architectures harness time-of-flight (TOF) information from spatially diverse reflection links, offering substantial improvements in localization accuracy \cite{godrich2010target, gogineni2011target, kanhere2021target}. However, fully leveraging both AOA and TOF measurements requires advanced signal-level designs that go beyond conventional spatial beamforming. This highlights a critical gap in current research: the systematic properties and ISAC trade-offs unique to D-ISAC systems remain underexplored, particularly in the context of transmit signal design.}

Building on these observations, we recognize that the potential performance gains from exploiting distributed configurations remain constrained by current design approaches. Therefore, it is imperative to investigate the achievable D-ISAC performance gains by incorporating optimal ISAC signal designs that jointly optimize sensing and communication performance. Motivated by this need, this work proposes a novel transmit signal design framework for D-ISAC systems. We consider \textcolor{black}{an orthogonal frequency division multiplexing (OFDM)-based} noncoherent D-ISAC system that operates without phase-level synchronization between nodes. While the performance gains of noncoherent cooperation may be limited compared to those of coherent systems \cite{yang2011phase}, \textcolor{black}{coherent processing requires accurate synchronization, which introduces excessive receiver complexity and signaling overheads}. Therefore, noncoherent systems are more practical for distributed wireless networks. The proposed system comprises multiple ISAC nodes connected to a central processing unit (CPU) and employs CoMP transmission for multi-user communication. For sensing, it leverages both collocated and distributed MIMO radar configurations with monostatic and multistatic target links, utilizing AOA and TOF estimation to enhance target localization. Notably, the proposed framework is not restricted to specific configurations and can be applied to systems employing distributed antenna arrays. Furthermore, low complexity of D-ISAC transmit signal design is crucial for practical implementation, particularly in distributed systems with large-scale configurations. Within this framework, we introduce a novel transmit signal design approach that balances the trade-offs between sensing and communication performance \textcolor{black}{together with implementation complexity} in noncoherent D-ISAC systems. The main contributions of this work are as follows:

\begin{itemize}
    \item 
    We propose a novel transmit signal design framework tailored for D-ISAC systems. This integrates both collocated and distributed MIMO radar for target localization and noncoherent CoMP downlink communication. For the sensing functionality, a hybrid localization approach is employed, leveraging both AOA and TOF measurements, which enables precise target localization by combining the benefits of TOF- and AOA-based techniques.
    
    \item 
    We develop a D-ISAC signal model and establish performance metrics for multi-target sensing and multi-user communication. Subsequently, optimization problems are formulated to design D-ISAC signals that minimize the target localization Cramér-Rao bound (CRB) while ensuring a minimum signal-to-interference-plus-noise ratio (SINR) for communication users. We employ semidefinite relaxation (SDR) to solve the formulated non-convex problems.
    
    \item 
    Three distinct transmit signal designs, including optimal, orthogonal, and beamforming designs, are introduced. The optimal design achieves the best sensing performance at the expense of computational complexity. The orthogonal design reduces computational requirements by ensuring inter-node signal orthogonality, while the beamforming design further simplifies implementation by focusing solely on optimizing AOA estimation performance. These designs are highly valuable for practical systems, offering flexible solutions that balance performance and complexity in D-ISAC systems.
\end{itemize}
Extensive numerical simulations validate the proposed signal designs, demonstrating D-ISAC trade-off performance bounds and assessing the impact of key system parameters, such as signal bandwidth, the number of antennas, and the number of cooperative nodes. Additionally, the analysis explores the performance-complexity trade-offs among the three proposed designs.

$Notations$: Boldface variables with lower- and upper-case symbols represent vectors and matrices, respectively. $\textbf{A} \in \mathbb{C}^{N \times M}$ and $\textbf{B} \in \mathbb{R}^{N \times M}$ denotes a complex-valued ${N \times M}$ matrix $\textbf{A}$ and a real-valued ${N \times M}$ matrix $\textbf{B}$, respectively. Also, $\mathbf{0}_{N \times M}$ and $\mathbf{I}_N$ denote a $N \times M$ zero-matrix and a $N \times N$ identity matrix, respectively. $({\cdot})^{T}$, $({\cdot})^{H}$, and $({\cdot})^{*}$ represent the transpose, Hermitian transpose, and conjugate operators, respectively. ${\text{diag}({\textbf{a}})}$ denotes a diagonal matrix with diagonal entries of a vector $\textbf{a}$. $\odot$, $\bullet$, $\ast$, and $\otimes$ represent the Hadamard (element-wise) product, the Khatri-Rao product, the face-splitting product, and the Kronecker product, respectively. $\mathbb{E}{[\cdot]}$ is the statistical expectation operator.

\section{System Model}

\begin{figure}[t!]
    \centering
    {\includegraphics[scale=0.45]{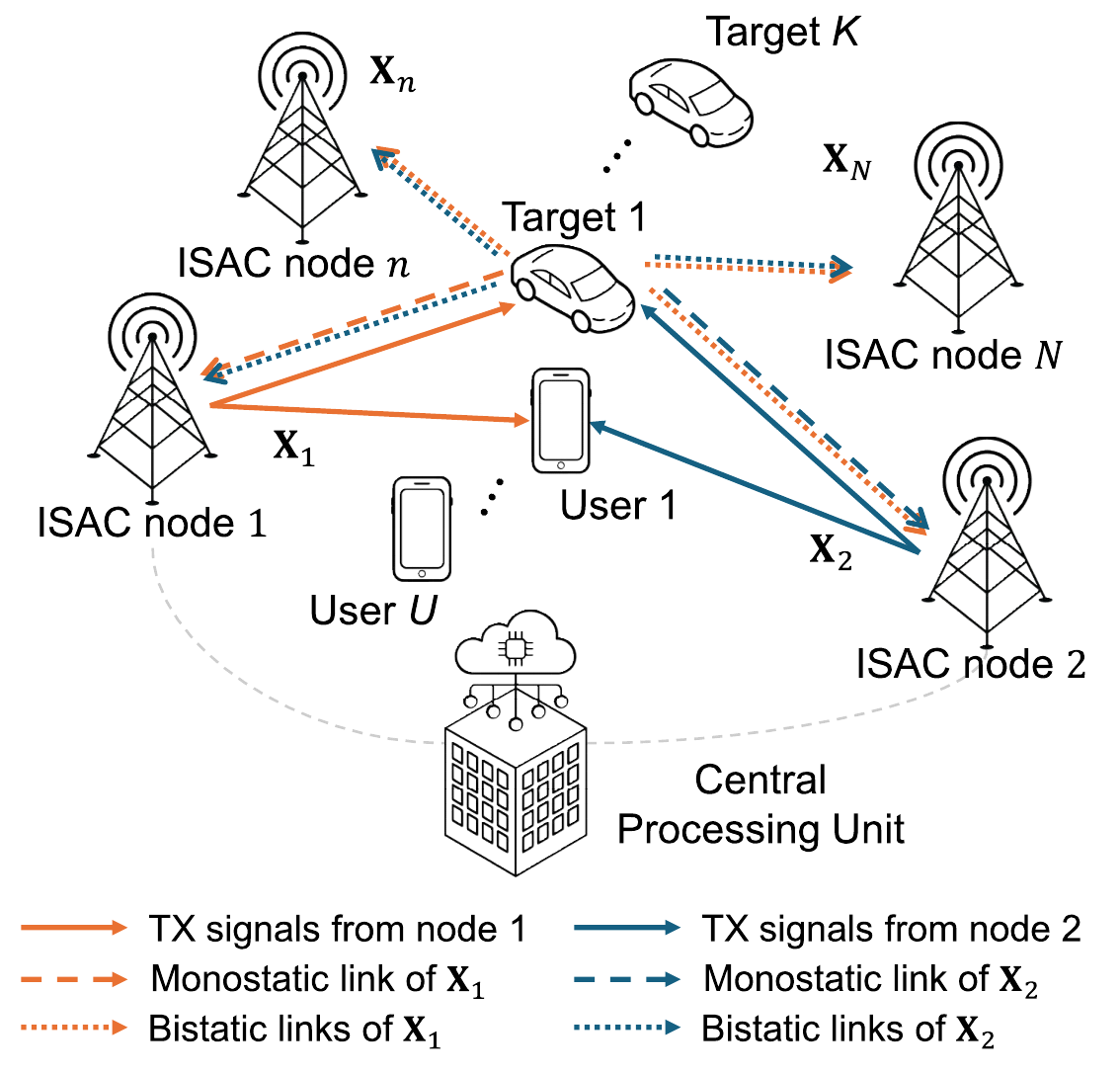}}
    \caption{Illustration of the D-ISAC system, where each ISAC node is equipped with both transmit and receive antenna arrays. Signals from other nodes are omitted for clarity. The D-ISAC system leverages all possible monostatic and bistatic links for target localization.}
    \label{f1}
\end{figure}

\subsection{Distributed ISAC System}
In the proposed D-ISAC system, we integrate noncoherent CoMP transmission for cooperative communication with a distributed MIMO radar framework that leverages both monostatic and bistatic sensing. \textcolor{black}{The noncoherent operation} introduces a novel trade-off between communication and sensing performance. Using a distributed MIMO radar to jointly exploit monostatic and bistatic scattering for target localization, the system improves the sensing performance with ISAC networks given a communication performance, as illustrated in Fig. \ref{f1}. To achieve this, all ISAC nodes are connected to a CPU, which is responsible for distributing communication data to each ISAC node and collecting radar-received signals for centralized fusion to estimate target parameters.

Each ISAC node is equipped with $M_t$ transmit antennas and $M_r$ receive antennas configured in a uniform linear array (ULA) with half-wavelength spacing. With $N$ ISAC nodes, node $n$ is located at $\mathbf{p}_n = [x_n, y_n]^T$ in a two-dimensional Cartesian coordinate system relative to the reference origin. Together, these networked ISAC nodes serve $U$ single-antenna communication users and jointly detect $K$ targets and estimate their corresponding parameters. This cooperative operation enables the distributed ISAC network to handle communication and sensing tasks simultaneously. The D-ISAC system employs OFDM with $L$ subcarriers over $T$ time slots.

For simplicity, we assume that the transmit and receive antennas are deployed separately with sufficient isolation, making the self-interference caused by full-duplex operation negligible. Furthermore, the transmitted signals and CSI between ISAC nodes are shared and known in advance. This prior knowledge facilitates the cancellation of inter-node interference in the received signals. To ensure practical implementation, we consider a noncoherent cooperation framework, in which the ISAC nodes are synchronized in time and frequency but not in phase. We note that time-frequency synchronization is relatively straightforward and can be achieved to some extent using global positioning system (GPS) disciplined references \cite{hofmann2012global}. On the other hand, phase synchronization is more difficult to achieve, because it requires aligning the instantaneous relative phases of distributed nodes that is highly sensitive to hardware-induced phase offset such as phase noises \cite{alemdar2021rfclock}.

\subsection{Transmit Signal Model}
\begin{table}[t!]
    \caption{Summary of D-ISAC Transmit Signal Designs}
    \centering
    \begin{tabular}{>{\centering\bfseries}m{0.6in} >{\centering}m{1in} >{\centering}m{0.65in} >{\centering\arraybackslash}m{0.62in}}
    \toprule
     & \textbf{D-ISAC Signal Model} & \textbf{Signal Correlation btw. Nodes} & 
    \textbf{Number of Variables} \\ 
    \midrule
    Optimal  (\(\mathcal{P}.\textit{1}\)) & $\mathbf{W}_{c,n,l}\mathbf{s}_{c,l} + \mathbf{x}_{s,n,l}$ & Non-orthogonal & $M_t^2N^2L + M_t^2NUL$ \\ \hline
    Orthogonal  (\(\mathcal{P}.\textit{2}\)) & $\mathbf{W}_{c,n,l}\mathbf{s}_{c,l} + \mathbf{x}_{s,n,l}$ & Orthogonal & $M_t^2(L + UL)$ \\ \hline
    Beamforming  (\(\mathcal{P}.\textit{3}\)) & $\mathbf{W}_{c,n,l}\mathbf{s}_{c,l} + \mathbf{W}_{s,n} \mathbf{s}_{s,l}$ & Orthogonal & $M_t^2(N + UL)$ \\ 
    \bottomrule
    \end{tabular}
    \label{tab::summary}
\end{table}

As a transmit signal model, we introduce a weighted summation of multi-user communication and MIMO radar signals as the ISAC transmit waveform. The frequency-domain OFDM ISAC signal at a symbol index $t \in [1, 2, \cdots, T]$ is expressed as
\begin{align}
    \mathbf{X}_n [t] = \mathbf{X}_{c,n} [t] + \mathbf{X}_{s,n} [t],   
    \label{Eqn::1}
\end{align}
where $\mathbf{X}_{c,n} [t] = \left[\mathbf{x}_{c,n,1} [t], \mathbf{x}_{c,n,2} [t], \cdots, \mathbf{x}_{c,n,L} [t]\right] \in \mathbb{C}^{M_t \times L}$, and the $l^{th}$ column vector of $\mathbf{X}_{c,n} [t]$, representing the $l^{th}$ subcarrier signal, is expressed as $\mathbf{x}_{c,n,l} [t] = \mathbf{W}_{c,n,l}\mathbf{s}_{c,l} [t] \in \mathbb{C}^{M_t \times 1}$. Here, $\mathbf{W}_{c,n,l} \in \mathbb{C}^{M_t \times U}$ is the precoding matrix for multi-user communication, and $\mathbf{s}_{c,l} [t] \in \mathbb{C}^{U \times 1}$ is the corresponding data symbol vector. Furthermore, $\mathbf{X}_{s,n} [t] \in \mathbb{C}^{M_t \times L}$ represents the radar signals at the ISAC node $n$. The power of communication and sensing signals is encapsulated in $\mathbf{X}_{c,n}[t]$ and $\mathbf{X}_{s,n}[t]$, respectively. Although a cyclic prefix (CP) is utilized to avoid inter-symbol interference (ISI), it is omitted in this formulation because the CP is removed during receiver processing at both the communication and radar receivers \cite{han2023sub}. Its effect is embedded into the OFDM symbol as a phase rotation.

Without loss of generality, we assume that the communication symbols are random with zero mean and uncorrelated with the radar signals, ensuring
\begin{align}
    \mathbb{E}\left[\mathbf{x}_{c,n,l}[t] \mathbf{x}^H_{s,m,l}[t]\right] = \mathbf{0}_{M_t \times M_t}, \quad \forall{n,m,} \; \text{and} \; l.
    \label{Eqn::2}
\end{align}
Additionally, the modulated communication symbols for different users are uncorrelated, i.e., 
\begin{align}
    \mathbb{E}\left[\mathbf{s}_{c,l}[t] \mathbf{s}^H_{c,l}[t]\right] =  \mathbf{I}_{U}, \quad \forall{l}.
    \label{Eqn::3}
\end{align}
For simplicity, we omit the symbol index $t$ in the following discussions. Based on this transmit signal model, the goal is to design the D-ISAC transmit signal $\mathbf{X}_n, \forall{n=1,2,\ldots,N}$ to meet the desired system performance requirements. Alternatively, this can be viewed as jointly designing the communication precoders $\mathbf{W}_{c,n}$ for multi-user CoMP transmission and the distributed MIMO radar signals $\mathbf{X}_{s,n}$, while balancing their respective signal powers to optimize the trade-off between sensing and communication performance.

Unlike an ISAC signal design in single-node systems \cite{liu2021cramer, liu2020joint}, where sensing performance is primarily determined by spatial beamforming for the radar waveform, the target localization performance of distributed MIMO radar is influenced by the ranging sequence across subcarriers and the signal correlation between different transmit nodes \cite{godrich2010target}. Consequently, we propose a comprehensive framework for D-ISAC transmitter design, incorporating subcarrier-level sensing signaling. By taking the signal correlation between nodes into account, we relax the optimal D-ISAC signal design to low-complexity orthogonal signal design. Furthermore, we present a block-level beamforming design, applying the same beamforming weights for all subcarriers, which further alleviates the design complexity compared to subcarrier-level design but achieves the suboptimal performance. The summary of three different types of designs are provided in Table \ref{tab::summary} and details of each signal model will be explored in Section \ref{main}.   

\subsection{Communication Signal Model}
According to the system description, each ISAC node transmits a dual-functional signal for radar sensing and downlink CoMP communication. Let $\mathbf{y}_{c,u,l}$ denote the received OFDM signal on the $l^{th}$ subcarrier at user $u$ in the frequency domain, which is expressed as
\begin{align}
   {y}_{c,u,l} & = \sum_{n = 1}^{N} \textbf{h}^{H}_{n,u,l} \mathbf{x}_{n,l} + {z}_{c,u,l}, \nonumber \\ 
   & = \sum_{n = 1}^{N} ( \textbf{h}^{H}_{n,u,l} \mathbf{w}_{c,n,u,l} {s}_{c,u,l} + \underbrace{\sum_{i = 1, i \neq u}^{U}  \textbf{h}^{H}_{n,u,l} \mathbf{w}_{c,n,i,l} {s}_{c,i,l}}_{\text{Multi-user interference}} ) \nonumber \\ 
   & \quad \quad \quad \quad \quad \quad + \underbrace{\sum_{n = 1}^{N}  \textbf{h}^{H}_{n,u,l} \mathbf{x}_{s,n,l}}_{\text{Radar interference}} + {z}_{c,u,l},
    \label{Eqn::4}
\end{align}

where $\textbf{h}^{H}_{n,u,l}$ represents the channel impulse response for the $l^{th}$ subcarrier between user $u$ and ISAC node $n$. The term ${z}_{c,u,l}$ represents zero-mean complex-valued additive white Gaussian noise (AWGN), following with ${z}_{c,u,l} \sim \mathcal{CN}(0,\sigma_c^2/L)$. In our system model, we consider an inter-carrier interference (ICI)-free scenario. As shown in (\ref{Eqn::2}), the received communication signal is impacted by multi-user interference as well as radar signal interference transmitted from all cooperative ISAC nodes. Therefore, the communication SINR for noncoherent CoMP accounts for these interference components.

\subsection{Sensing Signal Model}
For multiple ISAC nodes, the received signal at each node includes both monostatic and bistatic scatterings from $K$ targets. The received sensing signal at the $n^{\text{th}}$ ISAC node, transmitted from the $m^{\text{th}}$ node, is expressed as
\begin{equation}
    \mathbf{Y}_{s,n,m} = \sum_{k=1}^{K} b_{n,m}^k \mathbf{a}_r(\theta_n^k) \left( \mathbf{a}_t^T(\theta_m^k) \mathbf{X}_{m} \odot \mathbf{d}^T(\tau_{n,m}^k) \right),
    \label{Eqn::5}
\end{equation}   
where $\mathbf{a}_r(\theta) \in \mathbb{C}^{M_r \times 1}$ and $\mathbf{a}_t(\theta) \in \mathbb{C}^{M_t \times 1}$ are the receive and transmit steering vectors, respectively, and $\mathbf{d}(\tau) \in \mathbb{C}^{L \times 1}$ represents the delay steering vector. These are defined as:
\begin{subequations}\label{Eqn::6}
    \begin{align}
        \mathbf{a}_{r}(\theta) & = \left[1, e^{j\frac{2\pi}{\lambda} d_0 \sin{\theta}}, \ldots, e^{j\frac{2\pi}{\lambda} (M_r-1) d_0 \sin{\theta}} \right]^T, \label{Eqn::6a}\\
        \mathbf{a}_{t}(\theta) & = \left[1, e^{j\frac{2\pi}{\lambda} d_0 \sin{\theta}}, \ldots, e^{j\frac{2\pi}{\lambda} (M_t-1) d_0 \sin{\theta}} \right]^T, \label{Eqn::6b}\\
        \mathbf{d}(\tau) & = \left[ 1, e^{-j2\pi \frac{B}{L} \tau}, e^{-j2\pi \frac{B}{L} 2\tau}, \ldots, e^{-j2\pi \frac{B}{L} (L-1)\tau} \right]^T, \label{Eqn::6c}
    \end{align}
\end{subequations}
where $\lambda$ is the wavelength, $B$ is the signal bandwidth, and $L$ is the number of subcarriers. Additionally, $b_{n,m}^k$ is the complex amplitude including target radar cross-section (RCS) and path-loss corresponding to target-to-node distance, and $\tau_{n,m}^k = \tau_{n}^k + \tau_{m}^k$ is the time-of-flight (TOF) from the $m^{\text{th}}$ node to the $k^{\text{th}}$ target and then to the $n^{\text{th}}$ node. $\theta_n^k$ represents the angle of the target as seen from the $n^{\text{th}}$ node. Let $\mathbf{q}_k = [x_k, y_k]^T$ denote the position of target $k$. Then, the TOF and angle are functions of the target location, given by
\begin{subequations}\label{Eqn::7}
    \begin{align}
        \tau_{n}^k & =  \frac{\sqrt{(x_k - x_n)^2 + (y_k - y_n)^2}}{c}, \label{Eqn::7a}\\
        \theta_n^k & = \arctan2(y_k - y_n, x_k - x_n), \label{Eqn::7b}
    \end{align}
\end{subequations}
where $c$ is the speed of light.

From (\ref{Eqn::5}), the received signal at node $n$ is rewritten as
\begin{align}
    \mathbf{Y}_{s,n} & = \sum_{m=1}^{N} \left( \mathbf{A}_{r,n} \mathbf{B}_{n,m} \left( \mathbf{A}_{t,m}^T \mathbf{X}_m \odot \mathbf{D}_{n,m}^T \right) \right) + \mathbf{Z}_{s,n}, \nonumber \\
    & =  \mathbf{A}_{r,n} \sum_{m=1}^{N} \left( \mathbf{B}_{n,m} \mathbf{V}^{T}_{n,m}  \right) + \mathbf{Z}_{s,n},
    \label{Eqn::8}
\end{align}
where
\begin{subequations}\label{Eqn::9}
    \begin{align}
        \mathbf{V}_{n,m} & = \mathbf{X}^T_m \mathbf{A}_{t,m} \odot \mathbf{D}_{n,m}, \\
        \mathbf{A}_{r,n} & = \left[ \mathbf{a}_r(\theta_n^1), \mathbf{a}_r(\theta_n^2), \ldots, \mathbf{a}_r(\theta_n^K) \right],  \label{Eqn::10a} \\
        \mathbf{A}_{t,n} & = \left[ \mathbf{a}_t(\theta_n^1), \mathbf{a}_t(\theta_n^2), \ldots, \mathbf{a}_t(\theta_n^K) \right],  \label{Eqn::10b} \\
        \mathbf{B}_{n,m} & = \text{diag}(b_{n,m}^1, b_{n,m}^2, \ldots, b_{n,m}^K),  \label{Eqn::10c} \\
        \mathbf{D}_{n,m} & = \left[ \mathbf{d}(\tau_{n,m}^1), \mathbf{d}(\tau_{n,m}^2), \ldots, \mathbf{d}(\tau_{n,m}^K) \right], \label{Eqn::10d} \\
        \mathbf{D}_{n,m} & = \mathbf{D}_{n} \odot \mathbf{D}_{m},
    \end{align}
\end{subequations}
and $\mathbf{Z}_{s,n}$ represents circularly symmetric complex Gaussian noise, where each entry follows ${z}_{s,n,i,j} \sim \mathcal{CN}(0, \sigma_{n}^2), \forall{i} \in \{1, ... , M_r\}, \forall{j} \in \{1, ... , L\}$. For notational simplicity and without loss of generality, we assume all nodes have the same noise variance, $\sigma_{n}^2 = \sigma_s^2/L$. Consequently, the sensing task in the D-ISAC system is to estimate the target locations $\mathbf{q}_k$ using the collected signals from all nodes, expressed as
\begin{align}\label{Eqn::10}
    \mathbf{Y}_{s} = [\mathbf{Y}_{s,1}^T, \mathbf{Y}_{s,2}^T, \ldots, \mathbf{Y}_{s,N}^T]^T.
\end{align}

\subsection{Target Localization Methods in D-ISAC}
\begin{figure}[t!]
    \centering
    \subfloat[]{\includegraphics[scale=0.395]{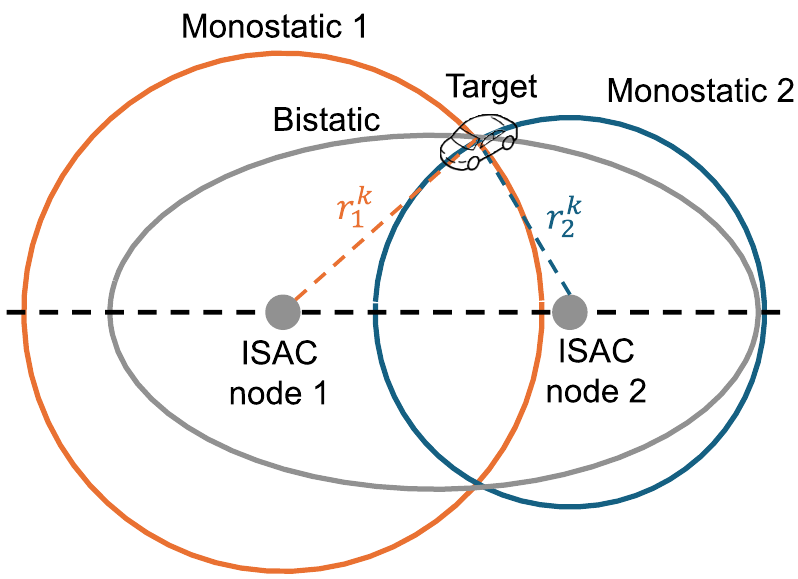}}
    \subfloat[]{\includegraphics[scale=0.395]{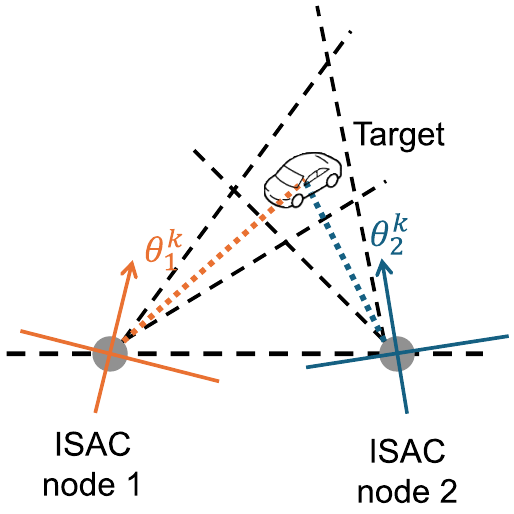}}
    \caption{Target localization methods in D-ISAC: (a) TOF-based localization and (b) AOA-based localization. The hybrid localization method combines TOF and AOA estimates to determine the target locations in the x-y coordinates.}
    \label{f2}
\end{figure}
From (\ref{Eqn::10}), we focus on estimating the x-y locations of targets as the sensing task. To achieve this, it is essential to determine which measurements are utilized during receiver processing. Fig. \ref{f2} illustrates two types of target localization methods in the proposed D-ISAC system, of which details are as follows.

\subsubsection{TOF-Based Localization}
TOF-based localization relies solely on TOF measurements, a widely adopted approach in distributed MIMO radar \cite{godrich2010target, sadeghi2021target}. In the proposed fully active D-ISAC system, both monostatic and bistatic target links are available, enabling the extraction of TOF information for all links, as shown in Fig. \ref{f2}(a). For this method, AOA information embedded in the signal model (\ref{Eqn::8}) is not utilized, making it equivalent to a distributed system with a single receive antenna per node or a system that performs TOF estimation after receiver beamforming toward the desired target direction. Consequently, angular information is not treated as an unknown parameter for estimation. This approach is particularly effective when the system operates with a large signal bandwidth, providing high-resolution TOF measurements for precise localization.

\subsubsection{AOA-Based Localization}
The D-ISAC system can also incorporate AOA information for target localization, as discussed in \cite{gao2023cooperative, zhao2020beamspace}. In the proposed full-duplex D-ISAC system, the angle-of-departure (AOD) at the \(n^{\text{th}}\) node is identical to its AOA. As shown in Fig. \ref{f2}(b), AOA-based localization estimates target angles relative to each node and translates them into x-y coordinates. This receiver processing is achieved by estimating AOA after range compression, which fixes the delay corresponding to the target range. As a result, the TOF information embedded in (\ref{Eqn::6c}) is not utilized for localization. Similar to TOF-based localization, AOA-based localization provides accurate results when the array aperture of each ISAC node is relatively large.

\subsubsection{Hybrid Localization}
The hybrid localization method leverages both TOF and AOA measurements to localize the target. This approach combines TOF- and AOA-based localization methods and is achieved by directly estimating the target locations \(\mathbf{q}_k\) from (\ref{Eqn::10}) without separating TOF and AOA measurements.

\begin{comment}
\begin{figure}[t!]
    \centering
    {\includegraphics[scale=0.5]{major_fig/Fig3.pdf}}
    \caption{.}
    \label{f3}
\end{figure}
\end{comment}

\section{Performance Metric for Distributed ISAC}
\subsection{Communication Performance Metric}
\label{comm_metric}
For noncoherent CoMP transmission, we exploit signal-to-interference-plus-noise ratio (SINR) as communication performance metric for D-ISAC signal design. From (\ref{Eqn::4}), we develop SINR of the typical user $u$ in the case of the noncoherent cooperation of the distributed ISAC system. Since the D-ISAC nodes are not phase-synchronized, the received signal at the typical communication user is a noncoherently combined signal of the jointly transmitted communication signals. The CP length has to be properly chosen to ensure ISI-free bistatic sensing between ISAC nodes, which also ensures an ISI-free CoMP transmission region for D-ISAC cooperation.

Accordingly, based on the received signal model, the SINR of the $l^{\text{th}}$ subcarrier of $u$ communication user can be developed as
\begin{equation}
    \gamma_{c,u,l} = \frac{\sum_{n = 1}^{N} |\textbf{h}^{H}_{n,u,l} \mathbf{w}_{c,n,u,l}|^2}{\sum_{n = 1}^{N} \left[\sum_{i = 1,i \neq u}^{U}  |\textbf{h}^{H}_{n,u,l} \mathbf{w}_{c,n,i,l}|^2  +  | \textbf{h}^{H}_{n,u,l} \mathbf{x}_{s,n,l}|^2\right] + \sigma_{c}^2/L}.
    \label{Eqn::SINR_NC}
\end{equation}
Note that the classical SINR expression in a coherent transmission scheme is represented as the square of the sum of all transmitted signals from $N$ nodes. This arises because the phase coherence among the transmitting nodes is preserved, resulting in the received signals being coherently combined. 

In contrast, for the noncoherent D-ISAC, the useful communication signals transmitted by the cooperative ISAC nodes are noncoherently combined at the user receiver, implying power combining of multiple transmitted signals. Also, it can be easily observed that the transmitted sensing signals and multi-user interference from all ISAC nodes are noncoherently added together, contributing the increase of interference power. Consequently, the SINR at the typical communication user is represented as the sum of the squared values of the transmitted signals from multiple ISAC nodes. One may refer to \cite{tanbourgi2014tractable} for the detailed proof of the SINR derivation.

\subsection{Sensing Performance Metric}
\label{sensing_metric}
For a sensing performance metric, we exploit CRB of target localization in the cooperative ISAC nodes. Let $\mathbf{\Psi}$ denote the collection of all the real-valued unknown parameters given as
\begin{equation}
    \mathbf{\Psi} = [\boldsymbol{\theta}^T, \boldsymbol{\tau}^T, \mathbf{b}_{\textsc{R}}^T, \mathbf{b}_{\textsc{I}}^T]^T \in \mathbb{R}^{(2KN+2KN^2) \times 1}, 
    \label{Eqn::12}
\end{equation} 
where
\begin{subequations}\label{Eqn::13}
    \begin{align}
        \label{Eqn::13(a)}
        \boldsymbol{\theta} & = [\theta_1^1, \theta_1^2, \ldots ,\theta_n^k, \ldots , \theta_N^K]^T \in \mathbb{R}^{KN \times 1} \\ \label{Eqn::13(b)}
        \boldsymbol{\tau} & = [\tau_1^1, \tau_1^2, \ldots ,\tau_n^k, \ldots , \tau_N^K]^T \in \mathbb{R}^{KN \times 1}  \\ \label{Eqn::13(c)}
        \mathbf{b} & = [b_{1,1}^1, b_{1,1}^2 , \ldots , b_{n,i}^k, \ldots, b_{N,N}^K]^T \in \mathbb{C}^{KN^2 \times 1},
    \end{align}
\end{subequations} 
and $\mathbf{b}_{\textsc{R}} = $\text{Re}$(\mathbf{b}) , \mathbf{b}_{\textsc{I}} = \text{Im}(\mathbf{b})$. Since both $\boldsymbol{\theta}$ and $\boldsymbol{\tau}$ contribute to localize targets in Cartesian coordination as (\ref{Eqn::7a}) and (\ref{Eqn::7b}), we can alternatively define the vector of unknown parameter as 
\begin{equation}
    \boldsymbol{\Theta} = [\mathbf{x}^T, \mathbf{y}^T, \mathbf{b}_{\textsc{R}}^T, \mathbf{b}_{\textsc{I}}^T]^T \in \mathbb{R}^{(2KN^2 + 2K) \times 1}, 
    \label{Eqn::11}
\end{equation} 
where
\begin{subequations}\label{Eqn::14}
    \begin{align}
        \mathbf{x} & = [x_1, x_2, \ldots , x_K]^T \in \mathbb{R}^{K \times 1} \\
        \mathbf{y} & = [y_1, y_2, \ldots , y_K]^T \in \mathbb{R}^{K \times 1}.
    \end{align}
\end{subequations} 

The joint log-likelihood function of the observation $\mathbf{Y}_s$ conditioned on $\mathbf{\Psi}$ is given by \cite{godrich2010target}
\begin{equation}
    \mathcal{L}(\mathbf{Y}_s | \mathbf{\Psi}) = -\sum_{n=1}^{N} \left\|\mathbf{Y}_{s,n} - \mathbf{A}_{r,n} \sum_{m=1}^{N} \left( \mathbf{B}_{n,m} \mathbf{V}^{T}_{n,m}  \right)\right\|_F^2 + c_{0},
    \label{Eqn::15}
\end{equation} 
where $c_{0}$ is some constant. Then, the Fisher Information matrix (FIM) with respect to $\mathbf{\Psi}$ is expressed as 
\begin{equation}
    \mathbf{F(\mathbf{\Psi})}  = \mathbb{E}\left[\left(\frac{\partial}{\partial{\mathbf{\Psi}}}\mathcal{L}(\mathbf{Y}_s | \mathbf{\Psi})\right)\left(\frac{\partial}{\partial{\mathbf{\Psi}}}\mathcal{L}(\mathbf{Y}_s | \mathbf{\Psi})\right)^T\right].
    \label{Eqn::16}
\end{equation} 
To obtain the FIM with respect to target localization parameters $\boldsymbol{\Theta}$, we adopt the chain rule as follows:
\begin{align}
    \mathbf{F(\boldsymbol{\Theta})}  & = \mathbf{J} \mathbf{F(\mathbf{\Psi})} \mathbf{J}^T, \\
    \mathbf{J} & = \frac{\partial{\mathbf{\Psi}}}{\partial{\boldsymbol{\Theta}}},
    \label{Eqn::17}
\end{align} 
where $\mathbf{J}$ is the Jacobian matrix. Finally, we get the sensing metric related to the target localization CRB, which is expressed as
\begin{equation}
    \text{CRB} = \text{tr} \left([\mathbf{F(\boldsymbol{\Theta})}]^{-1} \right).
    \label{Eqn::18}
\end{equation} 

\textcolor{black}{In the case of a coherent distributed MIMO radar system with the phase-level synchronization, one can leverage additional information corresponding to TOFs embedded in the phase of the complex amplitude $\mathbf{b}$, which enables to have coherent distributed MIMO radar gain for target localization \cite{godrich2010target}.
On the other hand, phase synchronization cannot be achieved between ISAC nodes for the noncoherent D-ISAC system, where the target complex amplitude $\mathbf{b}$ remains as nuisance parameters.} Therefore, (\ref{Eqn::12}), (\ref{Eqn::13}), and (\ref{Eqn::16}) are directly applied to get the FIM and the Jacobian matrix without any modification. The Jacobian matrix for the noncoherent combination is derived as
\begin{equation}
\mathbf{J} = 
    \begin{bmatrix}
        \frac{\partial \boldsymbol{\theta}}{\partial \mathbf{x}} & \frac{\partial \boldsymbol{\tau}}{\partial \mathbf{x}} & \mathbf{0} & \mathbf{0} \\
        \frac{\partial \boldsymbol{\theta}}{\partial \mathbf{y}} & \frac{\partial \boldsymbol{\tau}}{\partial \mathbf{y}} & \mathbf{0} & \mathbf{0} \\
        \mathbf{0} & \mathbf{0} & \mathbf{I}_{KN^2} & \mathbf{0} \\
        \mathbf{0} & \mathbf{0} & \mathbf{0} & \mathbf{I}_{KN^2} \\
    \end{bmatrix}.
    \label{Eqn::19}
\end{equation}
Each entry of the partial derivatives $\frac{\partial \boldsymbol{\theta}}{\partial \mathbf{x}}$, $\frac{\partial \boldsymbol{\theta}}{\partial \mathbf{y}}$, $\frac{\partial \boldsymbol{\tau}}{\partial \mathbf{x}}$, and $\frac{\partial \boldsymbol{\tau}}{\partial \mathbf{y}} \in \mathbb{R}^{K \times NK}$ is respectively obtained as

\begin{subequations}\label{Eqn::20}
    \begin{align}
        \frac{\partial \theta_n^k}{\partial x_k} & = \frac{y_n - y_k}{(x_k - x_n)^2 + (y_k - y_n)^2} \\
        \frac{\partial \theta_n^k}{\partial y_k} & = \frac{x_k - x_n}{(x_k - x_n)^2 + (y_k - y_n)^2} \\
        \frac{\partial \tau_n^k}{\partial x_k} & = \frac{x_k - x_n}{c \sqrt{(x_k - x_n)^2 + (y_k - y_n)^2}}\\
        \frac{\partial \tau_n^k}{\partial y_k} & = \frac{y_k - y_n}{c \sqrt{(x_k - x_n)^2 + (y_k - y_n)^2}}.
    \end{align}
\end{subequations} 
The detailed derivation of the FIM $\mathbf{F}(\mathbf{\Psi})$ for noncoherent D-ISAC is developed in Appendix \ref{FirstAppendix}.

\section{Distributed ISAC Signal Design} \label{main}
\subsection{Problem Formulation}
We propose a transmit signal design framework for a noncoherent D-ISAC system. Unlike monostatic ISAC systems, where transmit signal design primarily focuses on dual-functional radar-communication (DFRC) beamforming, the D-ISAC system requires a more advanced approach that considers signal design for optimal TOF estimation and inter-node signal correlation. \textcolor{black}{In D-ISAC, target localization performance depends on both TOF and AOA metrics, necessitating a transmit signal design that explicitly accounts for these parameters. This underscores the limitation of DFRC beamforming alone in achieving optimal localization performance for D-ISAC systems, as it overlooks the integration of TOF measurements and the power allocation across multiple transmit nodes.} To address these challenges, we develop a signal design framework that balances the trade-off between the target localization CRB and the CoMP communication SINR. Additionally, we propose three distinct D-ISAC signal design approaches: (1) optimal design, (2) orthogonal design, and (3) beamforming design. Each of these approaches demonstrates specific trade-offs between ISAC performance and computational complexity, allowing flexibility in system implementation depending on the performance requirements and resource constraints.

Using the communication and sensing performance metrics described in \ref{sensing_metric} and \ref{comm_metric}, the objective of the transmit signal design for noncoherent D-ISAC is to minimize the localization CRB while guaranteeing the communication SINR constraints for all downlink users. This can be generally formulated as the following optimization problem under a per-antenna power constraint:
\begin{subequations}\label{Eqn::P1-0}
    \begin{align}
    & \underset{\{\mathbf{X}_{n}\}}{\text{minimize}}
    & & \text{tr} \left([\mathbf{F(\boldsymbol{\Theta})}]^{-1} \right)  \label{Eqn::P1-0a} \\
    & \text{subject to}
    & & \left[\mathbf{R}_n\right]_{m,m}  \leq P_{T}, \; \forall{n, m}, \label{Eqn::P1-0b} \\
    & & & \gamma_{c,u,l} \geq \Gamma_{c}, \; \forall{u,l}, \label{Eqn::P1-0c}
    \end{align}
\end{subequations} 
where $\mathbf{R}_n = \mathbb{E}[\mathbf{X}_n\mathbf{X}_n^H]$ represents the transmit covariance matrix for the $n^{th}$ node. The objective function (\ref{Eqn::P1-0a}) minimizes the localization CRB, which reflects the target localization performance of the distributed MIMO radar. The constraint (\ref{Eqn::P1-0b}) imposes a per-antenna power constraint on all ISAC nodes, where $n$ and $m$ denote the node and antenna indices, respectively. The constraint (\ref{Eqn::P1-0c}) ensures that the SINR for each user across all OFDM subcarriers meets the threshold $\Gamma_{c}$, which allows the sensing operation without jeopardizing the communication quality of service (QoS). Since the optimization problem (\ref{Eqn::P1-0}) is a quadratic constrained quadratic problem (QCQP), we analyze and relax it into a convex problem using SDR. We explore three distinct signal designs and expand and tailor the above optimization to each of these, as described in the subsequent sections.

\subsection{Optimal Signal Design (\(\mathcal{P}.\textit{1}\))}
\label{Opt}
For the D-ISAC transmit signal design, we use SDR to solve problem (\ref{Eqn::P1-0}) without imposing any correlation constraint between ISAC signals transmitted from different nodes. A radar waveform utilizing subcarrier-by-subcarrier signaling, $\mathbf{X}_{s,n} = [\mathbf{x}_{s,n,1}, \dots, \mathbf{x}_{s,n,L}]$, is introduced, where each vector is defined as
\begin{equation}\label{Eqn::24}
    \mathbf{x}_{s,n,l} = \mathbf{W}_{s,n,l}\mathbf{s}_{s,l}, 
\end{equation}
with $\mathbf{W}_{s,n,l}$ as a $M_t \times M_t$ matrix and $\mathbf{s}_{s,l}$ as a $M_t \times 1$ radar signal column vector. Consistent with (\ref{Eqn::3}), we adopt uncorrelated radar sequences across transmit antennas in the same ISAC node, satisfying $\mathbb{E}[\mathbf{s}_{s,l} \mathbf{s}^H_{s,l}] =  \mathbf{I}_{M_t}, \forall{l} \in \{1, \dots , L\}$. In the proposed optimal D-ISAC signal design, since no constraints are imposed on ISAC signals between different nodes, there is no condition on $\mathbb{E}[\mathbf{x}_{n,l}\mathbf{x}_{m,l}^H], \forall{n \neq m}$. Cross-correlations of the signals between nodes are encapsulated in $\mathbf{W}_{n,l}\mathbf{W}_{m,l}^H$, where $\mathbf{W}_{n,l} = [\mathbf{W}_{c,n,l}, \mathbf{W}_{s,n,l}]$.

We first examine the FIM for target localization. By observing each block matrix derived in Appendix \ref{FirstAppendix}, each submatrix of $\mathbf{F}$ includes the products of $\mathbf{V}$, $\Dot{\mathbf{V}}_{\theta}$, and $\Dot{\mathbf{V}}_{\tau}$, with the node indices $n$ and $m$ omitted. The following lemma and theorem are then used to recast the quadratic constraint of problem (\ref{Eqn::P1-1}) into a linear expression.

\begin{lemma} \label{lemma1}
Any product of $\mathbf{V}$, $\Dot{\mathbf{V}}_{\theta}$, and $\Dot{\mathbf{V}}_{\tau}$ can be expressed in the following form:
\begin{equation}\label{eq::lemma1}
    \mathbf{V}_{n,i}^H \mathbf{V}_{n,j} = (\mathbf{A}_{t,i} \ast \mathbf{D}_{n,i})^H(\mathbf{X}_{i}^* \ast \mathbf{I}_L)(\mathbf{X}_{j} \ast \mathbf{I}_L)^T(\mathbf{A}_{t,j} \ast \mathbf{D}_{n,j}),
\end{equation}
\end{lemma}
\renewcommand\qedsymbol{$\blacksquare$}
\begin{proof}
    This result is readily derived using the face product property $(\mathbf{A}\mathbf{C})\odot(\mathbf{B}\mathbf{D}) = (\mathbf{A} \bullet \mathbf{B})(\mathbf{C} \ast \mathbf{D}) = (\mathbf{A}^T \ast \mathbf{B}^T)(\mathbf{C} \ast \mathbf{D})$ \cite{slyusar1999family}.
\end{proof}
\noindent Let $\tilde{\mathbf{x}}_{l}$ denote the $l^{\text{th}}$ column of $\tilde{\mathbf{X}}$, where $\tilde{\mathbf{X}} = \left[\mathbf{X}_{1}^T, \mathbf{X}_{2}^T, \dots, \mathbf{X}_{N}^T\right]^T$ is the augmented transmit signal matrix. By defining $\tilde{\mathbf{R}}_{l} = \mathbb{E}[\tilde{\mathbf{x}}_{l}\tilde{\mathbf{x}}^H_{l}]$, the following theorem provides insights for the FIM with respect to the D-ISAC transmit signal.
\begin{theorem} \label{theorem1}
The FIM $\mathbf{F}$ for target localization in the noncoherent D-ISAC is a linear function of $\tilde{\mathbf{R}}_{l}^T, \forall{l} \in \{1, 2, \dots, L\}$.
\end{theorem}

\renewcommand\qedsymbol{$\blacksquare$}
\begin{proof}
From Lemma \ref{lemma1}, $\mathbf{V}_{n,i}^H \mathbf{V}_{n,j}$ is a function of $(\mathbf{X}_i^* \ast \mathbf{I}_L)(\mathbf{X}_j \ast \mathbf{I}_L)^T$, which can be rewritten as
\begin{align}\label{eq::theorem1_(1)}
    (\mathbf{X}_i^* \ast \mathbf{I}_L)(\mathbf{X}_j \ast \mathbf{I}_L)^T & = \sum_{l = 1}^L (\mathbf{x}^*_{i,l} \otimes \mathbf{e}_l)(\mathbf{x}^T_{j,l} \otimes \mathbf{e}_l^T) \nonumber \\
    & = \sum_{l = 1}^L (\mathbf{x}^*_{i,l}\mathbf{x}^T_{j,l}) \otimes (\mathbf{e}_l\mathbf{e}_l^T).
\end{align}
Clearly, $\mathbb{E}[\mathbf{V}_{n,i}^H \mathbf{V}_{n,j}]$ is a linear function of $\mathbf{R}^T_{ij,l} = \mathbb{E}[\mathbf{x}^*_{i,l}\mathbf{x}^T_{j,l}], \forall{l} \in \{1, 2, \dots, L\}$. Based on (\ref{Eqn::2}), (\ref{Eqn::3}), and (\ref{Eqn::24}), we have 
\begin{align}\label{eq::theorem1_(2)}
    \mathbf{R}_{ij,l} = \mathbf{W}_{c,i,l}\mathbf{W}_{c,j,l}^H + \mathbf{W}_{s,i,l}\mathbf{W}_{s,j,l}^H.
\end{align}
Since the FIM includes $\mathbb{E}[\mathbf{V}_{n,i}^H \mathbf{V}_{n,j}], \forall{i,j} \in \{1, 2, \dots, N\}$, it is also a linear function of $\mathbf{R}^T_{ij,l}, \forall{i,j} \in \{1, 2, \dots, N\}, \; \text{and} \; \forall{l} \in \{1, 2, \dots, L\}$.

Defining the augmented communication precoding matrix $\tilde{\mathbf{W}}_{c,l} = [\mathbf{W}^T_{c,1,l}, \dots, \mathbf{W}^T_{c,N,l}]^T \in \mathbb{C}^{M_t N \times U}$, the augmented sensing signal matrix $\tilde{\mathbf{W}}_{s,l} = [\mathbf{W}^T_{s,1,l}, \dots, \mathbf{W}^T_{s,N,l}]^T \in \mathbb{C}^{M_t N \times M_t N}$, and $\tilde{\mathbf{W}}_{l} = [\tilde{\mathbf{W}}_{c,l}, \tilde{\mathbf{W}}_{s,l}] \in \mathbb{C}^{M_t N \times (U+M_t)}$ for the $l^{\text{th}}$ subcarrier, $\tilde{\mathbf{R}}_{l} = \mathbb{E}[\tilde{\mathbf{x}}_{l}\tilde{\mathbf{x}}^H_{l}]$ is expressed as
\begin{align}\label{eq::theorem1_(3)}
    \tilde{\mathbf{R}}_{l} & = \tilde{\mathbf{W}}_{c,l} \tilde{\mathbf{W}}_{c,l}^H + \tilde{\mathbf{W}}_{s,l} \tilde{\mathbf{W}}_{s,l}^H \nonumber \\
    & = \sum_{u=1}^{U} \tilde{\mathbf{w}}_{c,u,l}\tilde{\mathbf{w}}_{c,u,l}^H + \tilde{\mathbf{W}}_{s,l} \tilde{\mathbf{W}}_{s,l}^H, \nonumber \\
    & = \sum_{u=1}^{U} \tilde{\mathbf{R}}_{c,u,l} + \tilde{\mathbf{R}}_{s,l},
\end{align}
where $\tilde{\mathbf{w}}_{c,u,l}$ is the $u^{\text{th}}$ column of $\tilde{\mathbf{W}}_{c,l}$, $\tilde{\mathbf{R}}_{c,u,l} = \tilde{\mathbf{w}}_{c,u,l}\tilde{\mathbf{w}}_{c,u,l}^H$, and $\tilde{\mathbf{R}}_{s,l} = \tilde{\mathbf{W}}_{s,l} \tilde{\mathbf{W}}_{s,l}^H$. Equivalently, $\tilde{\mathbf{R}}_{l}$ can be represented as the following block matrix:
\begin{equation}\label{eq::theorem1_(4)}
    \tilde{\mathbf{R}}_{l} = \begin{bmatrix}
    {\mathbf{R}}_{11,l} & {\mathbf{R}}_{12,l} & \cdots & {\mathbf{R}}_{1N,l} \\
    {\mathbf{R}}_{21,l} & {\mathbf{R}}_{22,l} & \cdots & {\mathbf{R}}_{2N,l} \\
    \vdots & \vdots & \ddots & \vdots \\
    {\mathbf{R}}_{N1,l} & {\mathbf{R}}_{N2,l} & \cdots & {\mathbf{R}}_{NN,l}
    \end{bmatrix} \in \mathbb{C}^{M_{t}N \times M_{t}N}.
\end{equation}
Accordingly, the FIM is a linear function of $\tilde{\mathbf{R}}_{l}$, and this thus completes the proof.
\end{proof} 

\textbf{Remark 1:} Theorem 1 provides us some insights on the performance of the proposed D-ISAC system. Unlike a single-node ISAC transmit signal design, we can leverage hybrid of TOF and AOA information for the target localization in D-ISAC. Since the estimation of TOF is performed along the subcarrier-axis, i.e. $l = \{1,2, \cdots, L\}$, per-subcarrier sensing signal design is required to optimize the localization accuracy.    

We now rewrite the communication SINR for noncoherent CoMP in (\ref{Eqn::SINR_NC}) to relax the quadratic constraint (\ref{Eqn::P1-0c}). From (\ref{eq::theorem1_(2)}), it is readily found that 
\begin{align}\label{Eqn::31}
    \mathbf{R}_{nn,l} & = \sum_{u=1}^U  \mathbf{w}_{c,n,u,l}\mathbf{w}_{c,n,u,l}^H + \mathbf{W}_{s,n,l}\mathbf{W}_{s,n,l}^H \nonumber\\
    & = \sum_{u=1}^U \mathbf{R}_{c,n,u,l} + \mathbf{R}_{s,n,l},
\end{align}
where $\mathbf{R}_{c,n,u,l} = \mathbf{w}_{c,n,u,l}\mathbf{w}_{c,n,u,l}^H$ is rank-one. Using (\ref{Eqn::31}), the communication SINR in (\ref{Eqn::SINR_NC}) can be expressed as
\begin{equation}
    \gamma_{c,u,l} = \frac{\sum_{n = 1}^{N} \textbf{h}^{H}_{n,u,l} \mathbf{R}_{c,n,u,l} \textbf{h}_{n,u,l}}{\sum_{n = 1}^{N} \textbf{h}^{H}_{n,u,l} (\mathbf{R}_{nn,l} - \mathbf{R}_{c,n,u,l}) \textbf{h}_{n,u,l} + \sigma_{c}^2/L}.
    \label{Eqn::SINR_NC_re}
\end{equation}
Since $\mathbf{R}_{s,n,l}$ is encapsulated in $\mathbf{R}_{nn,l}$, the communication SINR is a linear constraint with respect to $\mathbf{R}_{nn,l}$ and $\mathbf{R}_{c,n,u,l}$. 

By adopting Theorem \ref{theorem1} with (\ref{eq::theorem1_(3)}) and substituting (\ref{Eqn::SINR_NC_re}) into the problem (\ref{Eqn::P1-0}), we reformulate the problem (\ref{Eqn::P1-0}) in a equivalent form using the Schur complement as follows \cite{zhang2006schur}:
\begin{subequations}\label{Eqn::P1-1}
    \begin{align}
    & \underset{\{\tilde{\mathbf{R}}_{c,u,l}\}, \{\tilde{\mathbf{R}}_{l}\}}{\text{minimize}}
    & & \sum_{i=1}^{2K} t_i \label{Eqn::P1-1a} \\
    & \text{subject to}
    & & \begin{bmatrix} \mathbf{F} & \mathbf{e}_i \\ \mathbf{e}^{T}_i & t_i \end{bmatrix} \succeq 0, \; \forall{i} \in \{1, \dots, 2K + 2KN^2\}, \label{Eqn::P1-1b}\\
    & & & [\mathbf{\tilde{R}}]_{m,m}  \leq P_{T}, \; \forall{m}, \label{Eqn::P1-1c} \\
    & & & \tilde{\mathbf{R}}_{c,u,l} \succeq 0, \; \text{rank}(\tilde{\mathbf{R}}_{c,u,l}) = 1, \forall{u,l}, \label{Eqn::P1-1d} \\
    & & & \tilde{\mathbf{R}}_l -\sum_{u=1}^{U}\tilde{\mathbf{R}}_{c,u,l} \succeq 0, \forall{u,l} \label{Eqn::P1-1e} \\
    & & &  \mkern-120mu \sum_{n = 1}^{N} \left[\left(1+\frac{1}{\Gamma_{c}}\right) \textbf{h}^{H}_{n,u,l} \mathbf{R}_{c,n,u,l} \textbf{h}_{n,u,l} - \textbf{h}^{H}_{n,u,l} \mathbf{R}_{nn,l} \textbf{h}_{n,u,l}\right] \geq \frac{\sigma_{c}^2}{L}, \label{Eqn::P1-1f}
    \end{align}
\end{subequations}
where $\mathbf{e}_i$ is the $i^{\text{th}}$ column of the identity matrix $\mathbf{I}_{(2K+2KN^2)}$, and $\mathbf{F}$ denotes the FIM matrix, simplified for brevity. Similar to (\ref{eq::theorem1_(4)}) where $\mathbf{R}_{nn,l}$ is a diagonal submatrix of $\tilde{\mathbf{R}}_l$, here $\mathbf{R}_{c,n,u,l}$ is also a diagonal submatrix of $\tilde{\mathbf{R}}_{c,u,l}$. The constraint (\ref{Eqn::P1-1c}) addresses the per-antenna power constraint, where $\forall{m} \in \{1, 2, \cdots, N M_t\}$ with $\mathbf{\tilde{R}} = \sum_{l = 1}^{L} \tilde{\mathbf{R}_l}$. The constraints (\ref{Eqn::P1-1d})--(\ref{Eqn::P1-1f}) hold $\forall{u} \in \{1, \cdots, U\}$ and $\forall{l} \in \{1, \cdots, L\}$. 

\textbf{Remark 2:} Here, one can observe that off-diagonal submatrices of $\tilde{\mathbf{R}}_{l}$ have no effects on the communication SINR constraint (\ref{Eqn::P1-1f}). This is because the noncoherent CoMP leverages a power combining gain without phase coherency, implying that the cross-correlation between different nodes does not contribute to the received signal power at a typical communication user.  Accordingly, we can simplify the problem by neglecting off-diagonal submatrics of $\tilde{\mathbf{R}}_{c,u,l}$. 

Dropping the rank-one constraint out, we recast (\ref{Eqn::P1-1}) as a semidefinite programming (SDP) problem:
\begin{subequations}\label{Eqn::P1-2}
    \begin{align}
    (\mathcal{P}.\textit{1})& \underset{\{\mathbf{R}_{c,n,u,l}\}, \{\tilde{\mathbf{R}}_{l}\}}{\text{minimize}}
    & & \sum_{i=1}^{2K} t_i  \label{Eqn::P1-2a} \\
    &  \text{subject to}
    & & \begin{bmatrix} \mathbf{F} & \mathbf{e}_i \\ \mathbf{e}^{T}_i & t_i \end{bmatrix} \succeq 0, \; \forall{i}, \label{Eqn::P1-2b}\\
    & & & [\mathbf{\tilde{R}}]_{m,m}  \leq P_{T}, \; \forall{m}, \label{Eqn::P1-2c} \\
    & & & \mathbf{R}_{c,n,u,l} \succeq 0, \;  \tilde{\mathbf{R}}_l \succeq 0, \forall{n,u,l}, \label{Eqn::P1-2d} \\
    & & & \mathbf{{R}}_{nn,l} - \sum_{u=1}^{U} \mathbf{R}_{c,n,u,l}\succeq 0, \forall{n,l}, \label{Eqn::P1-2e}\\
    & & &  \mkern-150mu \sum_{n = 1}^{N} \left[\left(1+\frac{1}{\Gamma_{c}}\right) \textbf{h}^{H}_{n,u,l} \mathbf{R}_{c,n,u,l} \textbf{h}_{n,u,l} - \textbf{h}^{H}_{n,u,l} \mathbf{R}_{nn,l} \textbf{h}_{n,u,l}\right] \geq \frac{\sigma_{c}^2}{L}. \label{Eqn::P1-2f}
    \end{align}
\end{subequations}
The formulated problem (\(\mathcal{P}.\textit{1}\)) is now a standard SDP, which can be solved via a CVX solver in polynomial time.

Although the relaxation to problem (\(\mathcal{P}.\textit{1}\)) is not tight, we can extract approximate solutions for $\hat{\mathbf{w}}_{c,n,u,l}$ and $\hat{\mathbf{W}}_{s,n,l}$ from the obtained solutions $\mathbf{R}_{c,n,u,l}$ and $\tilde{\mathbf{R}}_{l}$ of the relaxed problem. First, $\hat{\mathbf{w}}_{c,n,u,l}$ can be obtained from $\mathbf{R}_{c,n,u,l}$ as follows:
\begin{equation} 
    \hat{\mathbf{w}}_{c,n,u,l} = \left(\textbf{h}^{H}_{n,u,l} \mathbf{R}_{c,n,u,l} \textbf{h}_{n,u,l}\right)^{-1/2} \mathbf{R}_{c,n,u,l} \textbf{h}_{n,u,l}.
    \label{Eqn::34}
\end{equation}
Next, we construct the augmented communication precoding matrix $\widehat{\tilde{\mathbf{W}}}_{c,l}$ using the obtained $\hat{\mathbf{w}}_{c,n,u,l}$. Using (\ref{eq::theorem1_(3)}), the following covariance matrix of the radar signal can be constructed:
\begin{equation} 
   \widehat{\tilde{\mathbf{R}}}_{s,l} = \tilde{\mathbf{R}}_{l} - \widehat{\tilde{\mathbf{W}}}_{c,l}\widehat{\tilde{\mathbf{W}}}_{c,l}^H.
    \label{Eqn::35}
\end{equation}
However, $\widehat{\tilde{\mathbf{R}}}_{s,l}$ cannot be directly decomposed into $\widehat{\tilde{\mathbf{W}}}_{s,l}$. This is because the off-diagonal submatrices of $\widehat{\tilde{\mathbf{R}}}_{s,l}$ are not necessarily positive semidefinite, while the diagonal submatrices are positive semidefinite due to (\ref{Eqn::P1-2e}). Therefore, we first project $\widehat{\tilde{\mathbf{R}}}_{s,l}$ onto the positive semidefinite cone \cite{higham1988computing}, and then apply the Cholesky decomposition or eigenvalue decomposition \cite{higham1990analysis} to get
\begin{equation} 
   \widehat{\tilde{\mathbf{W}}}_{s,l}\widehat{\tilde{\mathbf{W}}}_{s,l}^H \approx \tilde{\mathbf{R}}_{l} - \widehat{\tilde{\mathbf{W}}}_{c,l}\widehat{\tilde{\mathbf{W}}}_{c,l}^H.
    \label{Eqn::36}
\end{equation}
The obtained solutions are not necessarily the optimal solution of the original problem (\ref{Eqn::P1-1}) because the relaxed problem does not account for the contribution of the communication precoding matrix to the off-diagonal matrices of $\tilde{\mathbf{R}}_{l}$. Nevertheless, these solutions can be used as feasible solutions for the optimal D-ISAC transmit signal design problem.

\subsection{Orthogonal Signal Design (\(\mathcal{P}.\textit{2}\))}
\label{Orth}
\begin{figure}[t!]
    \centering
    {\includegraphics[width=0.4\textwidth]{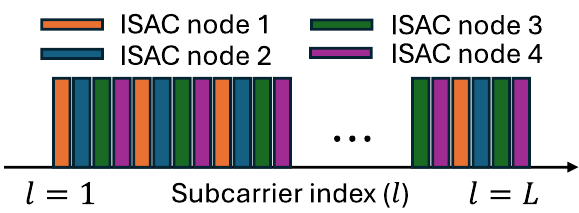}}
    \caption{Subcarrier-interleaving in OFDM D-ISAC signaling with orthogonal signal transmission when $N = 4$.}
    \label{f_sub}
\end{figure}
While the formulated problem (\(\mathcal{P}.\textit{1}\)) can provide convex relaxation bound for the optimal noncoherent D-ISAC performance, it incurs high computational complexity due to the large number of variables, which we analyze in Section IV-E. Importantly, the off-diagonal matrices of \(\tilde{\mathbf{R}}_{l}\) are not relevant to the communication SINR constraint in (\ref{Eqn::SINR_NC_re}). Leveraging these observations, we propose a suboptimal, low-complexity D-ISAC signal design based on orthogonal signal transmission across ISAC nodes.

To ensure the orthogonality of signals $\mathbf{x}_{n,l}$ and $\mathbf{x}_{m,l}$, the subcarrier-interleaving approach is adopted, where the subcarriers for each ISAC node are allocated such that they do not overlap as shown in Fig. \ref{f_sub}. Assuming the number of subcarriers is sufficiently larger than the number of ISAC nodes, i.e., $L \gg N$, the orthogonal signal model can be expressed as
\begin{align}\label{Eqn::37}
    \mathbf{x}_{n,l} & = 
    \begin{cases}
        \mathbf{W}_{c,n,l}\mathbf{s}_{c,l} + \mathbf{W}_{s,n,l}\mathbf{s}_{s,l}, & \text{if } n = l \; (\text{mod} \; N), \\
        \mathbf{0}_{M_t \times 1}, & \text{otherwise}.
    \end{cases}
\end{align}

This ensures that the cross-correlation between ISAC signals transmitted from different nodes is zero, irrespective of the sample number, i.e., $\mathbf{x}_{n,l}\mathbf{x}_{m,l}^H = \mathbf{0}_{M_t \times M_t}, \forall{n \neq m}$. By leveraging orthogonal signal transmission, the number of variables in the formulated problem can be significantly reduced compared to the case of the optimal signal design. This reduction is formalized in the following proposition. Hereafter, let us define the ISAC node index $\bar{n} = l \; (\text{mod} \; N)$, which serves the $l^{\text{th}}$ subcarrier.

\begin{proposition}
    For the D-ISAC transmit signal with orthogonal signals, the FIM for target localization is a linear function of $\mathbf{R}_{\bar{n}\bar{n},l}, \forall{(\bar{n},l)} \in \{(\bar{n},l) \;|\; \bar{n} = l \; (\text{mod} \; N), \forall{l} \in \{1, 2, \dots, L\}\}$.
\end{proposition}

\renewcommand\qedsymbol{$\blacksquare$}
\begin{proof}
    Based on (\ref{Eqn::37}), the ISAC signal of the $l^{\text{th}}$ subcarrier is only present in the transmitted signal of the ISAC node $\bar{n}$, where $\bar{n} = l \; (\text{mod} \; N)$. Consequently, all submatrices of $\tilde{\mathbf{R}}_{l}$ in (\ref{eq::theorem1_(4)}) are zero matrices, except for $\mathbf{R}_{\bar{n}\bar{n},l}$.
\end{proof}

We now reformulate the transmit signal design problem under the orthogonal signal condition based on Proposition 1. While the relaxation procedure remains the same as in the previous section, the variables are modified according to the orthogonal signal model. The communication SINR in (\ref{Eqn::SINR_NC_re}) is then updated as
\begin{equation}
    \gamma_{c,u,l} = \frac{ \textbf{h}^{H}_{\bar{n},u,l} \mathbf{R}_{c,\bar{n},u,l} \textbf{h}_{\bar{n},u,l}}{\textbf{h}^{H}_{\bar{n},u,l} (\mathbf{R}_{\bar{n}\bar{n},l} - \mathbf{R}_{c,\bar{n},u,l}) \textbf{h}_{\bar{n},u,l} + \sigma_{c}^2/L}.
    \label{Eqn::SINR_NC_re2}
\end{equation}

\noindent Applying Proposition 1 and the updated SINR expression in (\ref{Eqn::SINR_NC_re2}), and dropping the rank-one constraint in (\ref{Eqn::P1-1}), we formulate the relaxed problem for the D-ISAC transmit design with orthogonal signaling (\(\mathcal{P}.\textit{2}\)) as
\begin{subequations}\label{Eqn::P2-2}
    \begin{align}
    (\mathcal{P}.\textit{2})& \underset{\{\mathbf{R}_{c,\bar{n},u,l}\}, \{\mathbf{R}_{\bar{n}\bar{n},l}\}}{\text{minimize}}
    & & \sum_{i=1}^{2K} t_i  \label{Eqn::P2-2a} \\
    & \mkern+15mu \text{subject to}
    & & \begin{bmatrix} \mathbf{F} & \mathbf{e}_i \\ \mathbf{e}^{T}_i & t_i \end{bmatrix} \succeq 0, \; \forall{i}, \label{Eqn::P2-2b}\\
    & & & [\mathbf{\tilde{R}}]_{m,m}  \leq P_{T}, \; \forall{m}, \label{Eqn::P2-2c} \\
    & & & \mathbf{R}_{c,\bar{n},u,l} \succeq 0, \; \forall{\bar{n},u,l},  \label{Eqn::P2-2d} \\
    & & & \mathbf{R}_{\bar{n}\bar{n},l} - \sum_{u=1}^{U} \mathbf{R}_{c,\bar{n},u,l} \succeq 0, \; \forall{\bar{n},l},\label{Eqn::P2-2e}\\
    & & &  \mkern-180mu \left(1+\frac{1}{\Gamma_{c}}\right) \textbf{h}^{H}_{\bar{n},u,l} \mathbf{R}_{c,\bar{n},u,l} \textbf{h}_{\bar{n},u,l} - \textbf{h}^{H}_{\bar{n},u,l} \mathbf{R}_{\bar{n}\bar{n},l} \textbf{h}_{\bar{n},u,l} \geq \frac{\sigma_{c}^2}{L}, \label{Eqn::P2-2f}
    \end{align}
\end{subequations}
Here, the constraints (\ref{Eqn::P2-2d})--(\ref{Eqn::P2-2f}) hold for $\forall{(\bar{n},l)} \in \{(\bar{n},l) \;|\; \bar{n} = l \; (\text{mod}\; N), \forall{l} \in \{1, 2, \dots, L\}\}$ and $\forall{u} \in \{1, \dots, U\}$. The formulated problem (\(\mathcal{P}.\textit{2}\)) is a convex problem, which can be solved using a CVX solver. Compared to the optimal signal design problem (\(\mathcal{P}.\textit{1}\)), the number of variables in (\(\mathcal{P}.\textit{2}\)) is significantly reduced, which will be further analyzed in Section \ref{complexity}.

To extract the solution of $\hat{\mathbf{w}}_{c,\bar{n},u,l}$ and $\hat{\mathbf{W}}_{s,\bar{n},l}$, we exploit a similar approach as in (\ref{Eqn::34})--(\ref{Eqn::36}). Notably, the optimal solution of the original problem (\ref{Eqn::P1-0}) under orthogonal signal transmission can be derived from the optimal solution of $(\mathcal{P}.\textit{2})$, as supported by the following theorem.

\begin{theorem}
    Given that $\mathbf{R}_{c,\bar{n},u,l}$ and $\mathbf{R}_{\bar{n}\bar{n},l}$ are the optimal solutions of the relaxed problem $(\mathcal{P}.\textit{2})$, the following extracted solutions $\hat{\mathbf{w}}_{c,\bar{n},u,l}$ and $\hat{\mathbf{W}}_{s,\bar{n},l}$, expressed as
    \begin{align}
        & \hat{\mathbf{w}}_{c,\bar{n},u,l}  = \left(\textbf{h}^{H}_{\bar{n},u,l} \mathbf{R}_{c,\bar{n},u,l} \textbf{h}_{\bar{n},u,l}\right)^{-1/2} \mathbf{R}_{c,\bar{n},u,l} \textbf{h}_{\bar{n},u,l}, \label{Eqn::40}\\
        & \hat{\mathbf{W}}_{s,\bar{n},l}\hat{\mathbf{W}}_{s,\bar{n},l}^H  = \mathbf{R}_{\bar{n}\bar{n},l} - \hat{\mathbf{W}}_{c,\bar{n},l}\hat{\mathbf{W}}_{c,\bar{n},l}^H, \label{Eqn::41}
    \end{align} 
    constitute the optimal solution to (\ref{Eqn::P1-0}) under the orthogonal signal model (\ref{Eqn::37}).
\end{theorem}

\renewcommand\qedsymbol{$\blacksquare$}
\begin{proof}
    Since the $l^{\text{th}}$ OFDM subcarrier is served only by the ISAC node $\bar{n}$, the constraints (\ref{Eqn::P2-2d})--(\ref{Eqn::P2-2f}) are exclusively applied to ISAC node $\bar{n}$ for a given $l$. 
    Provided that $[\mathbf{R}_{\bar{n}\bar{n},l}]_{m,m} = \rho_{\bar{n},m,l} P_T$, where $\sum_{l=1}^{L} \rho_{\bar{n},m,l} \leq 1$, the optimal solution of $(\mathcal{P}.\textit{2})$ for each $l$ corresponds to that of a single-node ISAC transmit signal design. The remainder of the proof follows the same as in Appendix A of \cite{liu2020joint}.
\end{proof}
\noindent Therefore, the orthogonal signal transmission in D-ISAC, utilizing interleaved subcarriers, not only enables efficient D-ISAC transmit signal design but also ensures a rank-one optimal solution from the relaxed optimization problem.

%%$M_t(M_t+1)NUL/2 + M_tN(M_tN+1)L/2$.
%%According to Proposition 1, the number of variables for $\tilde{\mathbf{R}}_{l}$, $M_tN(M_tN+1)/2$, in problem (\(\mathcal{P}.\textit{1}\)) can be shrunk to $M_t(M_t+1)/2$.

\subsection{Beamforming Design (\(\mathcal{P}.\textit{3}\))}
\label{bf}
In this section, we focus on a simplified D-ISAC transmit signal design approach based on radar beamforming, diverging from the per-subcarrier signal design strategies discussed in Sections \ref{Opt} and \ref{Orth}. Instead of optimizing the radar waveform on a per-subcarrier basis, the beamforming design employs a unified spatial precoding structure for sensing signals across all subcarriers. This approach significantly reduces computational complexity while effectively utilizing the spatial degrees of freedom provided by the ISAC system. By emphasizing beamforming for radar waveforms, this method seeks to achieve a balance between system performance and implementation efficiency in D-ISAC signal design.

For the beamforming design in D-ISAC, we apply a block-level beamforming approach to the radar signal, where the signal model is expressed as
\begin{equation}\label{Eqn::42}
    \mathbf{x}_{s,n,l} = \mathbf{W}_{s,n}\mathbf{s}_{s,l}, 
\end{equation}
where \(\mathbf{W}_{s,n}\) represents the unified beamforming matrix for the radar signal. Compared to the signal model for the optimal signal design in (\ref{Eqn::24}), the beamforming design imposes the condition \(\mathbf{W}_{s,n} = \mathbf{W}_{s,n,1} = \cdots = \mathbf{W}_{s,n,L}\), ensuring the same beamforming weights across all subcarriers. 

Additionally, to simplify the design further, we incorporate the orthogonal signal condition given in (\ref{Eqn::37}). Then, the D-ISAC signal model for the beamforming design is constructed as:
\begin{align}\label{Eqn::43}
    \mathbf{x}_{n,l} & = 
    \begin{cases}
        \mathbf{W}_{c,n,l}\mathbf{s}_{c,l} + \mathbf{W}_{s,n}\mathbf{s}_{s,l}, & \text{if } n = l \; (\text{mod} \; N), \\
        \mathbf{0}_{M_t \times 1}, & \text{otherwise},
    \end{cases}
\end{align}
which allows us to leverage Proposition 1 to further reduce the design variables. The covariance matrix \(\mathbf{R}_{\bar{n}\bar{n},l}\) for the specific ISAC node \(\bar{n} = l \; (\text{mod} \; N)\) can then be rewritten as:
\begin{align}\label{Eqn::44}
    \mathbf{R}_{\bar{n}\bar{n},l} & = \sum_{u=1}^U \mathbf{R}_{c,\bar{n},u,l} + \mathbf{R}_{s,\bar{n}},
\end{align}
where \(\mathbf{R}_{s,\bar{n}}\) is the covariance matrix of the sensing signal for ISAC node \(\bar{n}\). 

\textbf{Remark 3:} Notably, we observe that the covariance matrix of the sensing signal is independent of the OFDM subcarriers with the signal model for beamforming design. This insight highlights a performance limit that designing beamforming for the sensing signal alone cannot optimize TOF estimates but contributes solely to angle-of-arrival AOA estimates for target localization.

Based on this observation, we can simplify the optimization problem by introducing the following averaged covariance matrix over the subcarriers:
\begin{align}\label{Eqn::45}
    \bar{\mathbf{R}}_{\bar{n}\bar{n}} & = \sum_{l=1}^{L} \sum_{u=1}^U \mathbf{R}_{c,\bar{n},u,l} + \frac{L}{N}\mathbf{R}_{s,\bar{n}},
\end{align}
where \(L\) is assumed to be a multiple of \(N\). By expressing the FIM as a function of the averaged covariance matrix \(\frac{N}{L}\bar{\mathbf{R}}_{\bar{n}\bar{n}}\), the transmit signal design contributes solely to AOA measurements in the target localization CRB. Although this relaxation does not provide the optimal solution for the D-ISAC transmit signal design, it reduces computational complexity by $L/N$ times compared to per-subcarrier sensing signal design approaches, which further will be discussed in Section \ref{complexity}.
 
Consequently, we ignore the dependency on the subcarrier in (\ref{Eqn::P2-2b}), leading to the relaxed optimization problem for the beamforming design as:
\begin{subequations}\label{Eqn::P3-2}
    \begin{align}
    (\mathcal{P}.\textit{3})& \underset{\{\mathbf{R}_{c,\bar{n},u,l}\}, \{\bar{\mathbf{R}}_{\bar{n}\bar{n}}\}}{\text{minimize}}
    & & \sum_{i=1}^{2K} t_i  \label{Eqn::P3-2a} \\
    & \mkern+15mu \text{subject to}
    & & \begin{bmatrix} \mathbf{F} & \mathbf{e}_i \\ \mathbf{e}^{T}_i & t_i \end{bmatrix} \succeq 0, \; \forall{i}, \label{Eqn::P3-2b}\\
    & & & [\mathbf{\tilde{R}}]_{m,m}  \leq P_{T}, \; \forall{m}, \label{Eqn::P3-2c} \\
    & & & \mathbf{R}_{c,\bar{n},u,l} \succeq 0, \; \forall{\bar{n},u,l},  \label{Eqn::P3-2d} \\
    & & & \bar{\mathbf{R}}_{\bar{n}\bar{n}} - \sum_{l=1}^{L} \sum_{u=1}^{U} \mathbf{R}_{c,\bar{n},u,l} \succeq 0, \; \forall{\bar{n}},\label{Eqn::P3-2e}\\
    & & &  \mkern-170mu \left(1+\frac{1}{\Gamma_{c}}\right) \textbf{h}^{H}_{\bar{n},u,l} \mathbf{R}_{c,\bar{n},u,l} \textbf{h}_{\bar{n},u,l} - \textbf{h}^{H}_{\bar{n},u,l} \mathbf{U}_{\bar{n}\bar{n},l} \textbf{h}_{\bar{n},u,l} \geq \frac{\sigma_{c}^2}{L}, \label{Eqn::P3-2f} \\
    & & & \mkern-130mu \mathbf{U}_{\bar{n}\bar{n},l} = \frac{N}{L}\left(\bar{\mathbf{R}}_{\bar{n}\bar{n}} - \sum_{l=1}^{L} \sum_{u=1}^{U} \mathbf{R}_{c,\bar{n},u,l}\right) + \sum_{u=1}^{U} \mathbf{R}_{c,\bar{n},u,l}. \label{Eqn::P3-2g}
    \end{align}
\end{subequations}

Although the FIM is evaluated using the averaged covariance matrix \(\bar{\mathbf{R}}_{\bar{n}\bar{n}}\), the per-subcarrier communication SINR constraint must still be preserved to ensure communication performance. Thus, the communication SINR constraint (\ref{Eqn::P2-2f}) in \((\mathcal{P}.\textit{2})\) is equivalently expressed as (\ref{Eqn::P3-2f}) and (\ref{Eqn::P3-2g}) using the variables \(\mathbf{R}_{c,\bar{n},u,l}\) and \(\bar{\mathbf{R}}_{\bar{n}\bar{n}}\). The formulated problem \((\mathcal{P}.\textit{3})\) is a standard SDP problem, which can be efficiently solved using a CVX solver.

After solving the convex problem, we obtain \(\hat{\mathbf{w}}_{c,\bar{n},u,l}\) using (\ref{Eqn::40}). Additionally, the extraction of \(\hat{\mathbf{W}}_{s,\bar{n}}\) is revised as:
\begin{align}
    \hat{\mathbf{W}}_{s,\bar{n}}\hat{\mathbf{W}}_{s,\bar{n}}^H  = \frac{N}{L}\left(\bar{\mathbf{R}}_{\bar{n}\bar{n}} - \sum_{l=1}^{L} \sum_{u=1}^{U} \mathbf{R}_{c,\bar{n},u,l}\right). \label{Eqn::47}
\end{align}
Referring to Theorem 2, the relaxation of the rank-one constraint from (\ref{Eqn::P1-1}) to (\(\mathcal{P}.\textit{3}\)) is tight for the orthogonal signal model as defined in (\ref{Eqn::43}). However, it is important to note that the proposed beamforming design in D-ISAC represents a suboptimal solution to (\ref{Eqn::P1-0}) because it solely optimizes the signaling for AOA measurements in target localization, without addressing TOF estimation.

\subsection{Computational Complexity Analysis} \label{complexity}

Aligned with the discussions of the three presented D-ISAC signal design approaches, we compare the computational complexity of solving the three different SDP problems: (\(\mathcal{P}.\textit{1}\)), (\(\mathcal{P}.\textit{2}\)), and (\(\mathcal{P}.\textit{3}\)). The time complexity of solving a general SDP problem with variable size \(n \times n\) and \(m\) constraints using an interior-point algorithm is given as \cite{jiang2020faster}:
\begin{align}
    C_T = \mathcal{O}(\sqrt{n}(mn^2 + m^w + n^w)\log(1/\epsilon)),
\end{align}
where \(w\) is the matrix multiplication exponent and \(\epsilon\) is the relative accuracy. Accordingly, the complexity is a polynomial function of the number of variables in the specific SDP problem.

Let us denote \(\mathcal{Q}_1\), \(\mathcal{Q}_2\), and \(\mathcal{Q}_3\) as the total number of variables for each problem (\(\mathcal{P}.\textit{1}\)), (\(\mathcal{P}.\textit{2}\)), and (\(\mathcal{P}.\textit{3}\)), respectively, which are computed as
\begin{align}
    \mathcal{Q}_1 & = M_t^2N^2L + M_t^2NUL, \label{Eqn::49}\\
    \mathcal{Q}_2 & = M_t^2(L + UL), \label{Eqn::50}\\
    \mathcal{Q}_3 & = M_t^2(N + UL). \label{Eqn::51}
\end{align}
Compared to the optimal signal design problem (\(\mathcal{P}.\textit{1}\)), the orthogonal signal transmission results in significantly lower complexity in (\(\mathcal{P}.\textit{2}\)), as these approaches do not consider signal cross-correlations between different nodes. The complexity of (\(\mathcal{P}.\textit{1}\)) increases substantially with the number of subcarriers \(L\). Furthermore, given that \(L \gg N\) where the number of subcarriers are much larger than the number of cooperative ISAC nodes, the beamforming design (\(\mathcal{P}.\textit{3}\)) exhibits even lower complexity since it focuses solely on optimizing AOA measurements for target localization and does not require subcarrier-by-subcarrier signal design.

\section{Numerical Simulation Results}
In this section, we present numerical simulation results for the proposed D-ISAC transmit signal design methods. The results focus on the performance trade-off between sensing and communication in the D-ISAC system using the proposed transmit signal designs. Additionally, we investigate the effects of various system parameters to provide insights into the TOF-AOA hybrid localization method and the three different types of D-ISAC signal designs. Unless state otherwise, the default system parameters are set as follows: \(P_T/\sigma_c^2 = P_T/\sigma_s^2 = 20\) dB, the number of transmit and receive antennas per ISAC node \(M = M_t = M_r = 6\), and the number of OFDM subcarriers \(L = 16\). The communication channel is modeled as a Rayleigh fading channel, with normalized channel gain to simplify the analysis for the multi-user scenario. The average target amplitude \(\sum_{k=1}^K\sum_{n=1}^N\sum_{m=1}^N |b_{n,m}^k|^2 / KN^2\) is fixed to \(10\) dB.

\subsection{Comparisons of Target Localization Methods}
\label{Simul1}
\begin{figure}[t!]
    \centering
    \subfloat[]{\includegraphics[width=0.4\textwidth]{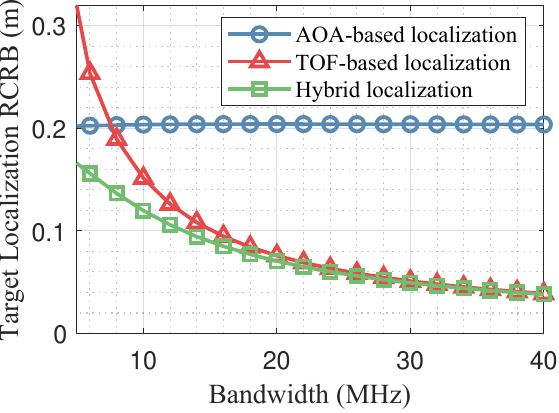}} \\
    \centering
    \subfloat[]{\includegraphics[width=0.4\textwidth]{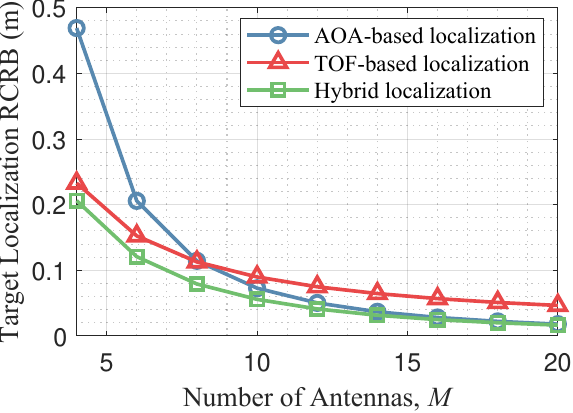}}
    \caption{Performance comparisons between AOA-based, TOF-based, and hybrid localization methods under varying D-ISAC system parameters: target localization RCRB vs. (a) signal bandwidth with \(M = 6\), and (b) the number of antennas, keeping the total power constant, with a \(10\)-MHz signal bandwidth. The designed D-ISAC signal with \(\Gamma_c = 30\) dB is used for evaluation with $K = 1, U = 1$.}
    \label{f4}
\end{figure}
We first examine the target localization methods for D-ISAC. The proposed D-ISAC system utilizes a hybrid TOF and AOA localization approach, leveraging all possible measurements to localize targets. This method combines the advantages of both collocated MIMO and distributed MIMO radar sensing. For this simulation, two ISAC nodes are positioned at \([35.35 \, \mathrm{m}, -35.35 \, \mathrm{m}]\) and \([-35.35 \, \mathrm{m}, -35.35 \, \mathrm{m}]\) with their antenna arrays oriented toward the origin. This corresponds to a 50 m distance from each node to the origin. A single user is located at \([0 \, \mathrm{m}, -20 \, \mathrm{m}]\), and a single target is positioned at the origin. To compare the target localization performance depending on localization methods, we employed the D-ISAC signaling with the orthogonal design (\(\mathcal{P}.\textit{2}\)). This choice was made because the sensing performance trends remain consistent across all signaling designs for the different localization methods. The CRB for the TOF-based localization method is evaluated by setting \(\Dot{\mathbf{A}}_{r,n}\) and \(\Dot{\mathbf{A}}_{t,n}\) to zero, while the CRB for the AOA-based localization method is evaluated by setting \(\Dot{\mathbf{D}}_{n}\) to zero in the FIM.

The results in Fig. \ref{f4} demonstrate the superior localization root-CRB (RCRB) of the hybrid localization method compared to the other two approaches. However, the performance trends among these methods vary depending on the system parameters. For instance, increasing the signal bandwidth primarily improves TOF measurements, as shown in Fig. \ref{f4}(a). This validates that the performance of the TOF-based localization method converges toward that of the hybrid approach with wideband D-ISAC systems. Conversely, increasing the number of antennas in the D-ISAC system significantly enhances the AOA-based localization performance compared to the TOF-based method in Fig. \ref{f4}(b). This is because the TOF-based localization benefits only from increased SNR owing to beamforming gain, whereas the AOA-based method further exploits the advantages of the larger array aperture size. These observations are closely linked to the ISAC performance comparisons among the proposed three different types of transmit signal designs, which will be further discussed in the following section.

\subsection{ISAC Performance of D-ISAC Transmit Signal Designs}
\label{ISAC_performance}
\begin{figure}[t!]
    \centering
    {\includegraphics[width=0.4\textwidth]{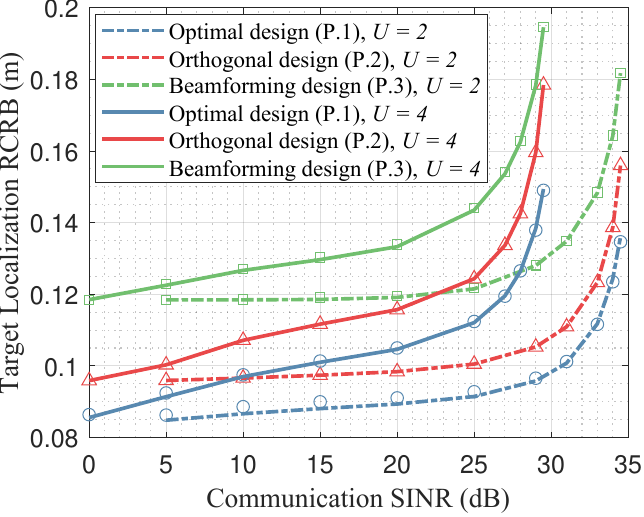}}
    \caption{ISAC trade-off performances of the proposed D-ISAC transmit signal designs with varying numbers of users. The number of targets is \(K = 2\). The markers (\(\textcolor{black}{{\Circle}}\), \(\textcolor{black}{\bigtriangleup}\), and \(\textcolor{black}{\square}\)) on the lines represent the extracted solutions derived from the optimal solutions of the respective SDP problems.}
    \label{f5}
\end{figure}

\begin{figure}[t!]
    \centering
    {\includegraphics[width=0.4\textwidth]{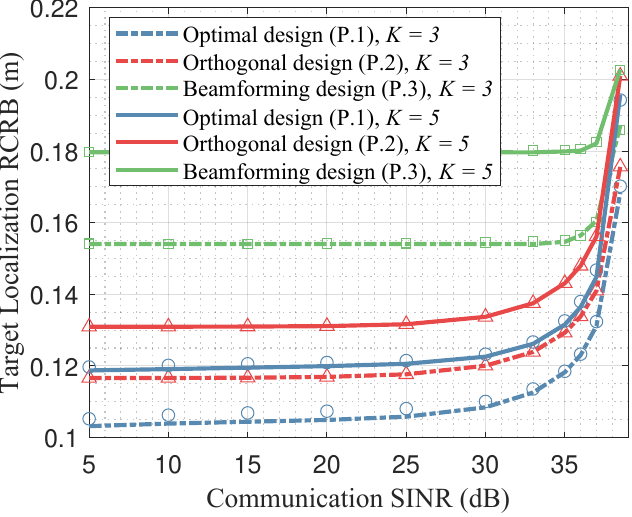}}
    \caption{ISAC trade-off performances of the proposed D-ISAC transmit signal designs with varying numbers of targets. The number of users is $U = 1$. The markers (\(\textcolor{black}{{\Circle}}\), \(\textcolor{black}{\bigtriangleup}\), and \(\textcolor{black}{\square}\)) on the lines represent the extracted solutions derived from the optimal solutions of the respective SDP problems.}
    \label{f6}
\end{figure}

In this section, we evaluate the ISAC trade-off performances among three different D-ISAC designs: optimal signal design (\(\mathcal{P}.\textit{1}\)), orthogonal design (\(\mathcal{P}.\textit{2}\)), and beamforming design (\(\mathcal{P}.\textit{3}\)). For the simulation setup, we consider multi-target and multi-user scenarios. Radar targets are positioned linearly along the x-axis between \(-30 \, \mathrm{m}\) and \(30 \, \mathrm{m}\), with \(y_k = 0, \forall{k} \in \{1, \dots, K\}\). Similarly, communication users are positioned linearly along the y-axis between \(-20 \, \mathrm{m}\) and \(20 \, \mathrm{m}\), with \(x_u = 0, \forall{u} \in \{1, \dots, U\}\). The number of ISAC nodes is \(N = 2\), consistent with Section \ref{Simul1}, and the OFDM signal bandwidth is set to \(20 \, \mathrm{MHz}\).

First, we validate the CRB-SINR trade-offs of the three designs, as shown in Fig. \ref{f5} and Fig. \ref{f6}. The proposed D-ISAC transmit signal designs clearly demonstrate trade-offs between sensing and communication performance. As expected, the optimal signal design (\(\mathcal{P}.\textit{1}\)) achieves the best sensing performance compared to the other two designs utilizing orthogonal signal transmission. This result indicates that orthogonal signals for distributed MIMO radar do not guarantee optimal sensing performance for target localization. 

Furthermore, the overall ISAC performance degrades as the number of users increases in all three designs, and the maximum achievable communication SINR also decreases, as shown in Fig. \ref{f5}. Conversely, an increase in the number of targets degrades the target localization RCRB, while the maximum achievable communication SINR remains unchanged, as shown in Fig. \ref{f6}. Interestingly, the performance gap trends between the three designs remain consistent regardless of the number of users or targets. This observation implies that the differences between the designs lie solely in the signal modeling and do not affect the degree-of-freedom (DoF) of the transmitter design.

As discussed in Theorem 2, the optimal solution of the relaxed problem with the orthogonal signal model is also the optimal solution of the original problem without the rank-one constraint relaxation. This is validated by the markers on the lines in each plot, which indicate the performance of the extracted solutions derived from the relaxed problem's solutions.  Notably, the extracted solution of (\(\mathcal{P}.\textit{1}\)), which serves as an approximate solution, does not deviate significantly from the convex optimization bound. However, it exhibits degraded performance when the communication SINR is set to relatively low values. This degradation arises because the rank of \(\widehat{\tilde{\mathbf{R}}}_{s,l}\) in (\ref{Eqn::35}) may become deficient due to the approximation to a positive semidefinite matrix.

\begin{figure}[t!]
    \centering
    {\includegraphics[width=0.4\textwidth]{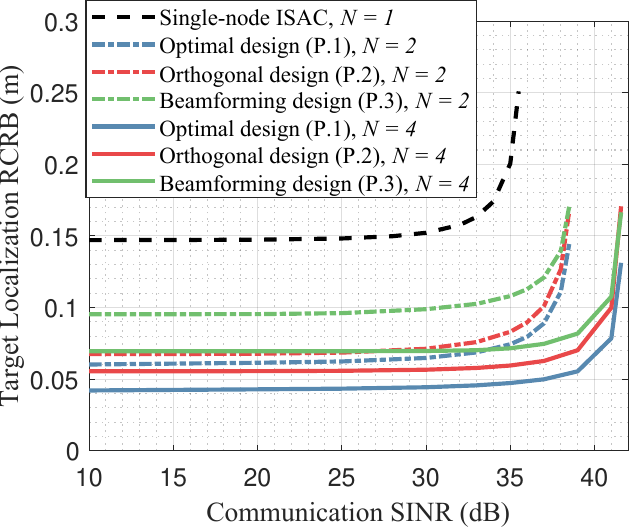}}
    \caption{ISAC trade-off performances of the proposed D-ISAC transmit signal designs with varying numbers of ISAC nodes. The number of targets and users is $K = 1, U = 1$. Each node is equipped with $M = 6$.}
    \label{f7}
\end{figure}

In Fig. \ref{f7}, we illustrate the ISAC performance gain of D-ISAC with varying numbers of cooperative nodes. For the case of \(N = 4\), the ISAC nodes are positioned at \([-46.19 \, \mathrm{m}, -19.13 \, \mathrm{m}]\), \([46.19 \, \mathrm{m}, -19.13 \, \mathrm{m}]\), \([-19.13 \, \mathrm{m}, -46.19 \, \mathrm{m}]\), and \([19.13 \, \mathrm{m}, -46.19 \, \mathrm{m}]\), all equidistant from the origin. As the number of cooperative nodes increases, we observe performance improvements in both sensing and communication. From the perspective of communication performance, this cooperation gain stems from noncoherent CoMP transmission, which leverages the power combining gain of distributed ISAC nodes. For target localization, the performance gain arises from the noncoherent processing of the distributed MIMO radar, utilizing TOF measurements to localize the target \cite{godrich2010target}. It is worth noting that this cooperative sensing gain is significantly influenced by the geometry of the target and node positions \cite{sadeghi2021target}. Furthermore, the optimal number of distributed ISAC nodes can be determined by considering the antenna-to-node allocation for a given cooperative node density \cite{meng2024network}.

\begin{figure}[t!]
    \centering
    {\includegraphics[width=0.475\textwidth]{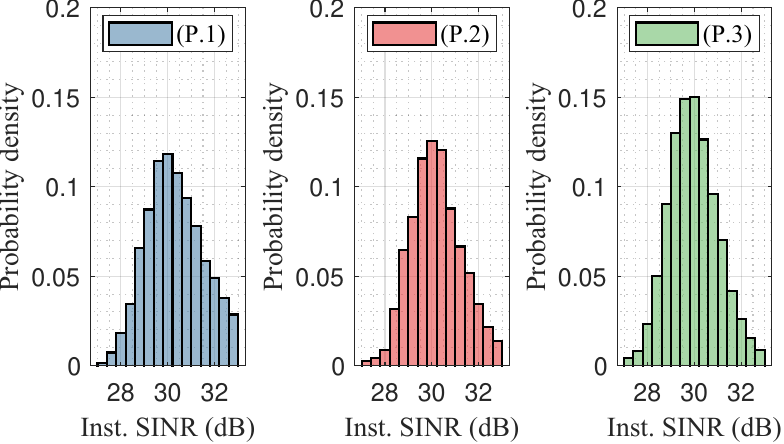}}
    \caption{Instantaneous communication SINR distributions for three different designs (\(\mathcal{P}.\textit{1}\)), (\(\mathcal{P}.\textit{2}\)), and (\(\mathcal{P}.\textit{3}\)) with $U = 2, K=2,$ and $\Gamma_c = 30$ dB.}
    \label{f8}
\end{figure}

\begin{figure}[t!]
    \centering
    {\includegraphics[width=0.4\textwidth]{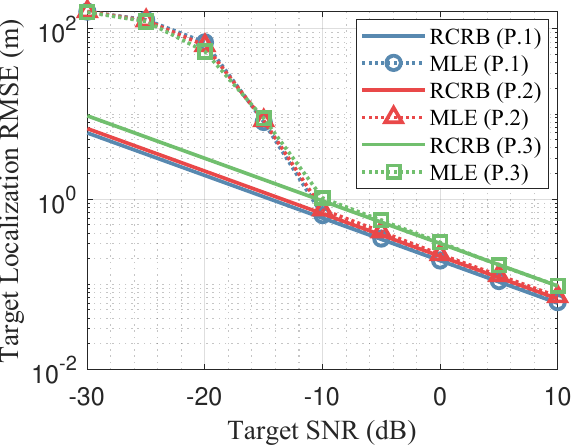}}
    \caption{Target localization performance with three different designs (\(\mathcal{P}.\textit{1}\)), (\(\mathcal{P}.\textit{2}\)), and (\(\mathcal{P}.\textit{3}\)) with $U = 1, K=1,$ and $\Gamma_c = 10$ dB.}
    \label{f9}
\end{figure}

We also evaluate the instantaneous communication SINR and target localization performance to validate the feasibility of the proposed D-ISAC transmit signal design based on the CRB-SINR trade-off. For this evaluation, we perform a Monte Carlo simulation with 1000 trials. The probability density functions (PDFs) of the instantaneous SINR based on the actual transmitted symbols for the three different designs are shown in Fig. \ref{f8}. The instantaneous SINR is primarily distributed between 28 dB and 32 dB for all three designs, with similar variances. This consistency arises from the use of average SINR as the communication performance metric, which does not guarantee symbol-level performance for instantaneous SINR. Fig. \ref{f9} illustrates the target localization root-mean square error (RMSE) compared to the CRB obtained from the proposed D-ISAC designs. The target location is estimated using a maximum likelihood estimator (MLE) for noncoherent distributed MIMO radar. The performance trends among the three designs align with the ISAC trade-off results, particularly for high-SNR targets, demonstrating consistency between the CRB and RMSE.

\subsection{Effects of System Parameters}

\begin{figure}[t!]
    \centering
    {\includegraphics[width=0.4\textwidth]{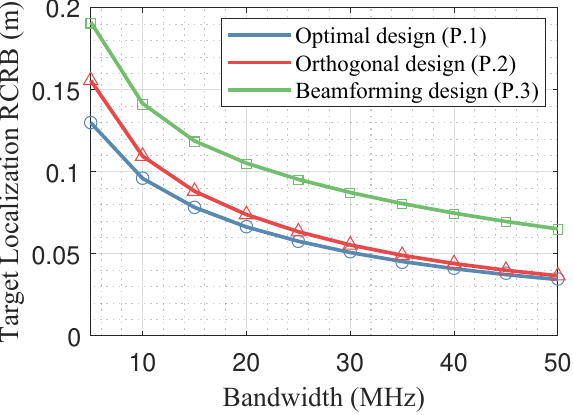}}
    \caption{Performance comparisons of three different designs (\(\mathcal{P}.\textit{1}\)), (\(\mathcal{P}.\textit{2}\)), and (\(\mathcal{P}.\textit{3}\)) with \(U = 1\), \(K = 1\), and \(\Gamma_c = 30\) dB under varying signal bandwidth with \(M = 6\). The markers (\(\textcolor{black}{{\Circle}}\), \(\textcolor{black}{\bigtriangleup}\), and \(\textcolor{black}{\square}\)) on the lines represent the extracted solutions derived from the optimal solutions of the respective SDP problems.}
    \label{f10}
\end{figure}

\begin{figure}[t!]
    \centering
    {\includegraphics[width=0.4\textwidth]{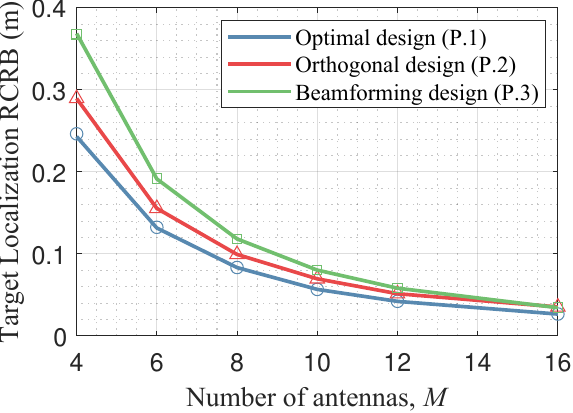}}
    \caption{Performance comparisons of three different designs (\(\mathcal{P}.\textit{1}\)), (\(\mathcal{P}.\textit{2}\)), and (\(\mathcal{P}.\textit{3}\)) with \(U = 1\), \(K = 1\), and \(\Gamma_c = 30\) dB under varying numbers of antennas, keeping the total power constant, with a \(5\)-MHz signal bandwidth. The markers (\(\textcolor{black}{{\Circle}}\), \(\textcolor{black}{\bigtriangleup}\), and \(\textcolor{black}{\square}\)) on the lines represent the extracted solutions derived from the optimal solutions of the respective SDP problems.}
    \label{f11}
\end{figure}

We explore the effects of D-ISAC system parameters on the performance of the proposed transmit signal designs. Here, we showcase the performance variations with respect to the signal bandwidth and the number of antennas. In Fig. \ref{f10}, the target localization CRBs as a function of the signal bandwidth are depicted. Interestingly, the performance of the orthogonal design (\(\mathcal{P}.\textit{2}\)) converges to that of the optimal design (\(\mathcal{P}.\textit{1}\)) as the signal bandwidth increases. This phenomenon can be explained as follows: TOF measurements predominantly contribute to target localization performance in wideband systems, as discussed in Section \ref{Simul1}. Since both (\(\mathcal{P}.\textit{1}\)) and (\(\mathcal{P}.\textit{2}\)) employ per-subcarrier signal designs optimized for TOF estimation, their performances become similar with increasing signal bandwidth. The primary difference between the two lies in the consideration of signal cross-correlation between ISAC nodes, which is more relevant to AOA estimation performance.

Conversely, as the number of antennas increases, the performance of the orthogonal design (\(\mathcal{P}.\textit{2}\)) converges to that of the beamforming design (\(\mathcal{P}.\textit{3}\)). Unlike the case of increasing signal bandwidth, a larger number of antennas enhances AOA estimation performance, and the hybrid localization performance becomes dominated by AOA measurements. Consequently, the per-subcarrier signal designs in (\(\mathcal{P}.\textit{1}\)) and (\(\mathcal{P}.\textit{2}\)) become less effective when AOA measurements predominantly determine target localization performance.

\subsection{Computational Complexity}
\begin{figure}[t!]
    \centering
    {\includegraphics[width=0.4\textwidth]{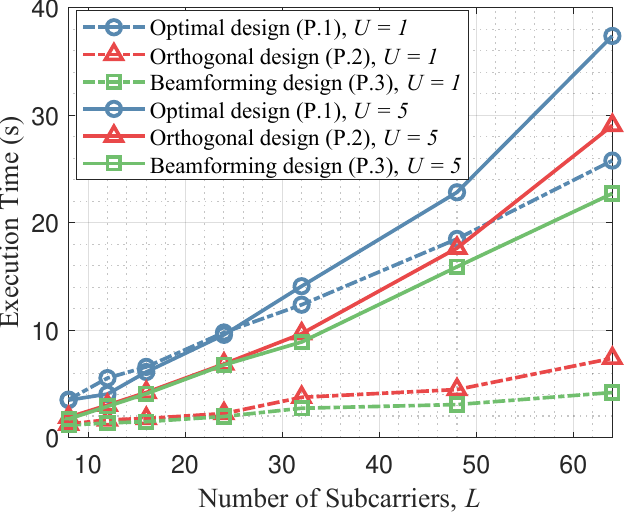}}
    \caption{Comparisons of the execution time with respect to the number of subcarriers with \(K = 1\).}
    \label{f12}
\end{figure}
Followed by the computational complexity analysis in Section \ref{complexity}, we empirically evaluate the execution time of each D-ISAC transmit signal design. The measured execution times are averaged over 100 trials. For all three design methods, the execution time increases with the number of subcarriers, consistent with the complexity trends discussed in (\ref{Eqn::49})--(\ref{Eqn::51}). As expected, the optimal signal design (\(\mathcal{P}.\textit{1}\)) exhibits the longest execution time among the three designs. Combined with the ISAC performances described in Section \ref{ISAC_performance}, the three designs demonstrate clear performance-complexity trade-offs. Consequently, one can select an appropriate design based on the system requirements such as computational resources.

\section{Conclusions}
This work presents a D-ISAC transmit signal design framework that enhances radar and communication functionalities through the cooperation of distributed ISAC systems. By leveraging both collocated and distributed MIMO radar, the proposed framework integrates AOA and TOF measurements for precise target localization, while improving communication SINR via the power combining gain of CoMP transmission. CRB-SINR-based optimization problems are formulated, and three different transmit signal designs—optimal, orthogonal, and beamforming—are introduced, highlighting trade-offs between ISAC performance and computational complexity. Numerical simulations validate the effectiveness of the proposed designs, providing insights into D-ISAC performance bounds, system parameter impacts, and performance-complexity trade-offs. The results demonstrate that the orthogonal design significantly reduces computational complexity while maintaining suboptimal performance, and the beamforming design achieves further complexity reductions with decreased performance trade-offs compared to the optimal design. These findings emphasize the importance of advanced transmit signal designs in unlocking the full potential of D-ISAC systems for future wireless networks. Future work will explore system-level aspects such as time-frequency synchronization, phase coherency, and ISAC receiver processing, which may significantly influence overall D-ISAC performance.

\appendices
\section{Derivation of the FIM for D-ISAC target localization}
\label{FirstAppendix}
Based on (\ref{Eqn::16}), we derive the FIM $\mathbf{F} (\mathbf{\Psi})$ for noncoherent distributed ISAC systems. For simplicity of the derivation, $\mathbf{F} (\mathbf{\Psi})$ can be expressed as a following block matrix:

\begin{equation}
    \mathbf{F} (\mathbf{\Psi}) = \frac{2}{\sigma^2} \text{Re} \left\{ \mathbb{E} \left(\begin{bmatrix} \mathbf{T} & \mathbf{E} \\ \mathbf{E}^T & \mathbf{G} \end{bmatrix}\right) \right\},
    \label{Eqn::ap_nc1}
\end{equation}
where $\mathbf{T} \in \mathbb{C}^{2KN \times 2KN}$, $\mathbf{E} \in \mathbb{C}^{2KN \times 2KN^2}$, and $\mathbf{G} \in \mathbb{C}^{2KN^2 \times 2KN^2}$. First, the submatrix $\mathbf{T}$ is also represented as
\begin{equation}
    \mathbf{T} =  \begin{bmatrix} \mathbf{F_{\boldsymbol{\theta}\boldsymbol{\theta}}} & \mathbf{F_{\boldsymbol{\theta}\boldsymbol{\tau}}} \\ \mathbf{F}_{\boldsymbol{\theta}\boldsymbol{\tau}}^{T} & \mathbf{F_{\boldsymbol{\tau}\boldsymbol{\tau}}} \end{bmatrix}.
    \label{Eqn::ap_nc2}
\end{equation}
We redefine (\ref{Eqn::13(a)}) and (\ref{Eqn::13(b)}) as $\boldsymbol{\theta}  = [\boldsymbol{\theta}_1^T, \boldsymbol{\theta}_2^T, \ldots ,\boldsymbol{\theta}_N^T]^T$ and $\boldsymbol{\tau}  = [\boldsymbol{\tau}_1^T, \boldsymbol{\tau}_2^T, \ldots ,\boldsymbol{\tau}_N^T]^T$, where $\boldsymbol{\theta}_n  = [\theta_n^1, \theta_n^2, \ldots ,\theta_n^k]^T$ and $\boldsymbol{\tau}_n  = [\tau_n^1, \tau_n^2, \ldots ,\tau_n^k]^T$ for $n = 1, 2, ..., N$. Also, $\mathbf{F_{\boldsymbol{\theta}\boldsymbol{\theta}}}, \mathbf{F_{\boldsymbol{\theta}\boldsymbol{\tau}}}$ , and $\mathbf{F_{\boldsymbol{\tau}\boldsymbol{\tau}}}$ are again partitioned into the block matrix with $K \times K$ submatrices denoted as $\mathbf{F_{\boldsymbol{\theta}_\textit{n}\boldsymbol{\theta}_\textit{m}}}, \mathbf{F_{\boldsymbol{\theta}_\textit{n}\boldsymbol{\tau}_\textit{m}}}$ , and $\mathbf{F_{\boldsymbol{\tau}_\textit{n}\boldsymbol{\tau}_\textit{m}}}$ for $n = 1, 2, ..., N$ and $m = 1, 2, ..., N$. Then,
each submatrix in (\ref{Eqn::ap_nc2}) can be developed by (\ref{Eqn::16}) as

\begin{subequations}
    \begin{align}
        \left\{\mathbf{F_{\boldsymbol{\theta}_\textit{n}\boldsymbol{\theta}_\textit{m}}}\right\}_{n = m} & = (\Dot{\mathbf{A}}_{r,n}^H \Dot{\mathbf{A}}_{r,n}) \odot (\mathbf{P}_{n}^H \mathbf{P}_{n}) \nonumber \\
        & \quad \quad \quad \quad + (\Dot{\mathbf{A}}_{r,n}^H \mathbf{A}_{r,n}) \odot (\mathbf{P}_{n}^H \Dot{\mathbf{V}}_{\theta_n,n,n} \mathbf{B}_{n,n}) \nonumber \\
        & \quad \quad \quad \quad + (\mathbf{A}_{r,n}^H \Dot{\mathbf{A}}_{r,n}) \odot (\mathbf{B}_{n,n}^{*} \Dot{\mathbf{V}}_{\theta_n,n,n}^H \mathbf{P}_{n}) \nonumber \\
        &  \mkern-10mu + \sum_{i=1}^N (\mathbf{A}_{r,i}^H \mathbf{A}_{r,i}) \odot (\mathbf{B}_{i,n}^{*} \Dot{\mathbf{V}}_{\theta_n,i,n}^H \Dot{\mathbf{V}}_{\theta_n,i,n} \mathbf{B}_{i,n}), \\
        \left\{\mathbf{F_{\boldsymbol{\theta}_\textit{n}\boldsymbol{\theta}_\textit{m}}}\right\}_{n \neq m} & = (\Dot{\mathbf{A}}_{r,n}^H \mathbf{A}_{r,n}) \odot (\mathbf{P}_{n}^H \Dot{\mathbf{V}}_{\theta_m,n,m} \mathbf{B}_{n,m}) \nonumber \\
        & \quad \quad \quad \quad + (\mathbf{A}_{r,m}^H \Dot{\mathbf{A}}_{r,m}) \odot (\mathbf{B}_{m,n}^{*} \Dot{\mathbf{V}}_{\theta_n,m,n}^H \mathbf{P}_{m}) \nonumber \\
        &  \mkern-20mu + \sum_{i=1}^N (\mathbf{A}_{r,i}^H \mathbf{A}_{r,i}) \odot (\mathbf{B}_{i,n}^{*} \Dot{\mathbf{V}}_{\theta_n,i,n}^H \Dot{\mathbf{V}}_{\theta_m,i,m} \mathbf{B}_{i,m}), \\
        \left\{\mathbf{F_{\boldsymbol{\theta}_\textit{n}\boldsymbol{\tau}_\textit{m}}}\right\}_{n = m} & = (\Dot{\mathbf{A}}_{r,n}^H \mathbf{A}_{r,n}) \odot (\mathbf{P}_{n}^H \Dot{\mathbf{P}}_{\tau_n,n}) \nonumber \\
        & \quad \quad \quad \quad + (\Dot{\mathbf{A}}_{r,n}^H \mathbf{A}_{r,n}) \odot (\mathbf{P}_{n}^H \Dot{\mathbf{V}}_{\tau_n,n,n} \mathbf{B}_{n,n}) \nonumber \\
        & \quad \quad \quad \quad + (\mathbf{A}_{r,n}^H {\mathbf{A}}_{r,n}) \odot (\mathbf{B}_{n,n}^{*} \Dot{\mathbf{V}}_{\theta_n,n,n}^H \Dot{\mathbf{P}}_{\tau_n,n}) \nonumber \\
        &  \mkern-10mu + \sum_{i=1}^N (\mathbf{A}_{r,i}^H \mathbf{A}_{r,i}) \odot (\mathbf{B}_{i,n}^{*} \Dot{\mathbf{V}}_{\theta_n,i,n}^H \Dot{\mathbf{V}}_{\tau_n,i,n} \mathbf{B}_{i,n}), \\
        \left\{\mathbf{F_{\boldsymbol{\theta}_\textit{n}\boldsymbol{\tau}_\textit{m}}}\right\}_{n \neq m} & = (\Dot{\mathbf{A}}_{r,n}^H \mathbf{A}_{r,n}) \odot (\mathbf{P}_{n}^H \Dot{\mathbf{V}}_{\tau_m,n,m} \mathbf{B}_{n,m}) \nonumber \\
        & \quad \quad \quad \quad + (\mathbf{A}_{r,m}^H {\mathbf{A}}_{r,m}) \odot (\mathbf{B}_{m,n}^{*} \Dot{\mathbf{V}}_{\theta_n,m,n}^H \Dot{\mathbf{P}}_{\tau_m,m}) \nonumber \\
        &  \mkern-20mu + \sum_{i=1}^N (\mathbf{A}_{r,i}^H \mathbf{A}_{r,i}) \odot (\mathbf{B}_{i,n}^{*} \Dot{\mathbf{V}}_{\theta_n,i,n}^H \Dot{\mathbf{V}}_{\tau_m,i,m} \mathbf{B}_{i,m}),\\
        \left\{\mathbf{F_{\boldsymbol{\tau}_\textit{n}\boldsymbol{\tau}_\textit{m}}}\right\}_{n = m} & = ({\mathbf{A}}^H_{r,n} \mathbf{A}_{r,n}) \odot (\Dot{\mathbf{P}}_{\tau_n,n}^H \Dot{\mathbf{P}}_{\tau_n,n}) \nonumber \\
        & \quad \quad \quad \quad + (\mathbf{A}_{r,n}^H {\mathbf{A}}_{r,n}) \odot (\Dot{\mathbf{P}}_{\tau_n,n}^H \Dot{\mathbf{V}}_{\tau_n,n,n} \mathbf{B}_{n,n} ) \nonumber \\
        & \quad \quad \quad \quad + (\mathbf{A}_{r,n}^H {\mathbf{A}}_{r,n}) \odot (\mathbf{B}_{n,n}^{*} \Dot{\mathbf{V}}_{\tau_n,n,n}^H \Dot{\mathbf{P}}_{\tau_n,n}) \nonumber \\
        &  \mkern-10mu + \sum_{i=1}^N (\mathbf{A}_{r,i}^H \mathbf{A}_{r,i}) \odot (\mathbf{B}_{i,n}^{*} \Dot{\mathbf{V}}_{\tau_n,i,n}^H \Dot{\mathbf{V}}_{\tau_n,i,n} \mathbf{B}_{i,n}), \\
        \left\{\mathbf{F_{\boldsymbol{\tau}_\textit{n}\boldsymbol{\tau}_\textit{m}}}\right\}_{n \neq m} & = (\mathbf{A}_{r,n}^H {\mathbf{A}}_{r,n}) \odot (\Dot{\mathbf{P}}_{\tau_n,n}^H \Dot{\mathbf{V}}_{\tau_m,n,m} \mathbf{B}_{n,m} ) \nonumber \\
        & \quad \quad \quad \quad + (\mathbf{A}_{r,m}^H {\mathbf{A}}_{r,m}) \odot (\mathbf{B}_{m,n}^{*} \Dot{\mathbf{V}}_{\tau_n,m,n}^H \Dot{\mathbf{P}}_{\tau_m,m}) \nonumber \\
        &  \mkern-20mu + \sum_{i=1}^N (\mathbf{A}_{r,i}^H \mathbf{A}_{r,i}) \odot (\mathbf{B}_{i,n}^{*} \Dot{\mathbf{V}}_{\tau_n,i,n}^H \Dot{\mathbf{V}}_{\tau_m,i,m} \mathbf{B}_{i,m}),
    \end{align}
    \label{Eqn::ap_nc3}
\end{subequations}
where $\mathbf{F_{\boldsymbol{\theta}_\textit{n}\boldsymbol{\theta}_\textit{m}}}$ is the $(m,n)$ submatrix of $\mathbf{F_{\boldsymbol{\theta}\boldsymbol{\theta}}}$. All derivatives with respect to unknown variables and related expressions are defined as
\begin{subequations}
    \begin{align}
        \mathbf{P}_{n} & = \sum_{m=1}^{N} \left( \mathbf{B}_{n,m} \mathbf{V}^{T}_{n,m}  \right)^T, \\
        \Dot{\mathbf{P}}_{\tau_n,n} & = \sum_{m=1}^{N} \left( \mathbf{B}_{n,m} \Dot{\mathbf{V}}^{T}_{\tau_n,n,m}  \right)^T, \\
        \Dot{\mathbf{V}}_{\theta_m,n,m} & = \mathbf{X}^T_m \Dot{\mathbf{A}}_{t,m} \odot \mathbf{D}_{n,m}, \\
        \Dot{\mathbf{V}}_{\tau_n,n,m} & = \Dot{\mathbf{V}}_{\tau_m,n,m} = \mathbf{X}^T_m \mathbf{A}_{t,m} \odot \Dot{\mathbf{D}}_{n,m}, \\
        \Dot{\mathbf{A}}_{r,n} & = \begin{bmatrix} \frac{\partial \mathbf{a}_r(\theta_n^1)}{\partial \theta_n^1}, \cdots, \frac{\partial \mathbf{a}_r(\theta_n^K)}{\partial \theta_n^K} \end{bmatrix}, \\
        \Dot{\mathbf{A}}_{t,n} & = \begin{bmatrix} \frac{\partial \mathbf{a}_t(\theta_n^1)}{\partial \theta_n^1}, \cdots, \frac{\partial \mathbf{a}_t(\theta_n^K)}{\partial \theta_n^K} \end{bmatrix}, \\
        \Dot{\mathbf{D}}_{n} & = \begin{bmatrix} \frac{\partial \mathbf{d}(\tau_{n}^1)}{\partial \tau_n^1}, \cdots, \frac{\partial \mathbf{d}(\tau_{n}^K)}{\partial \tau_n^K} \end{bmatrix},\\
        \Dot{\mathbf{D}}_{n,m} & = \Dot{\mathbf{D}}_{n} \odot {\mathbf{D}}_{m} = {\mathbf{D}}_{n} \odot \Dot{\mathbf{D}}_{m}.
    \end{align}
    \label{Eqn::ap_nc4}
\end{subequations}

Similar to (\ref{Eqn::ap_nc2}), $\mathbf{E}$ and $\mathbf{G}$ also can be represented as
\begin{subequations}
    \begin{align}
    \label{Eqn::ap_nc5(a)}
    \mathbf{E} & =  \begin{bmatrix} \mathbf{F_{\boldsymbol{\theta}\mathbf{b}}} & j\mathbf{F_{\boldsymbol{\theta}\mathbf{b}}} \\ \mathbf{F_{\boldsymbol{\tau}\mathbf{b}}} & j\mathbf{F_{\boldsymbol{\tau}\mathbf{b}}} \end{bmatrix}, \\ \label{Eqn::ap_nc5(b)}
    \mathbf{G} & =  \begin{bmatrix} \mathbf{F_{\mathbf{b}\mathbf{b}}} & j\mathbf{F_{\mathbf{b}\mathbf{b}}} \\ j\mathbf{F^{\textit{T}}_{\mathbf{b}\mathbf{b}}} & \mathbf{F_{\mathbf{b}\mathbf{b}}} \end{bmatrix}.
    \end{align}
    \label{Eqn::ap_nc5}
\end{subequations}
Let us redefine (\ref{Eqn::13(c)}) as $\mathbf{b}  = [\mathbf{b}_1^T, \mathbf{b}_2^T, \ldots ,\mathbf{b}_N^T]^T$, where $\mathbf{b}_n  = [\mathbf{b}_{n,1}^T, \mathbf{b}_{n,2}^T, \ldots ,\mathbf{b}_{n,N}^T]^T$ and $\mathbf{b}_{n,m}  = [{b}_{n,m}^1, {b}_{n,m}^2, \ldots ,{b}_{n,m}^K]^T$. By using the same notation with (\ref{Eqn::ap_nc3}), each submatrix in (\ref{Eqn::ap_nc5(a)}) is derived as follows:
\begin{subequations}
    \begin{align}
        \mathbf{F}_{\boldsymbol{\theta}_\textit{n}\mathbf{b}_{m}} & = \left[\mathbf{F}_{\boldsymbol{\theta}_\textit{n}\mathbf{b}_{m,1}},\mathbf{F}_{\boldsymbol{\theta}_\textit{n}\mathbf{b}_{m,2}},\ldots, \mathbf{F}_{\boldsymbol{\theta}_\textit{n}\mathbf{b}_{m,N}}\right], \\
        \mathbf{F}_{\boldsymbol{\tau}_\textit{n}\mathbf{b}_{m}} & = \left[\mathbf{F}_{\boldsymbol{\tau}_\textit{n}\mathbf{b}_{m,1}},\mathbf{F}_{\boldsymbol{\tau}_\textit{n}\mathbf{b}_{m,2}},\ldots, \mathbf{F}_{\boldsymbol{\tau}_\textit{n}\mathbf{b}_{m,N}}\right],
    \end{align}
    \label{Eqn::ap_nc6}
\end{subequations}
where
\begin{subequations}
    \begin{align}
        \left\{\mathbf{F}_{\boldsymbol{\theta}_\textit{n}\mathbf{b}_{m,i}}\right\}_{n=m} & = (\Dot{\mathbf{A}}_{r,n}^H \mathbf{A}_{r,n}) \odot (\mathbf{P}_{n}^H \mathbf{V}_{n,i}) \nonumber \\
        & \quad \quad \quad \quad + (\mathbf{A}_{r,n}^H \mathbf{A}_{r,n}) \odot (\mathbf{B}_{n,n}^{*} \Dot{\mathbf{V}}_{\theta_n,n,n}^H \mathbf{V}_{n,i}) \\ 
        \left\{\mathbf{F}_{\boldsymbol{\theta}_\textit{n}\mathbf{b}_{m,i}}\right\}_{n \neq m} & = (\mathbf{A}_{r,m}^H \mathbf{A}_{r,m}) \odot (\mathbf{B}_{m,n}^{*} \Dot{\mathbf{V}}_{\theta_n,m,n}^H \mathbf{V}_{m,i}),\\
        \left\{\mathbf{F}_{\boldsymbol{\tau}_\textit{n}\mathbf{b}_{m,i}}\right\}_{n= m} & =  (\mathbf{A}_{r,n}^H \mathbf{A}_{r,n}) \odot (\Dot{\mathbf{P}}_{\tau_n,n}^H \mathbf{V}_{n,i}) \nonumber \\
        & \quad \quad \quad \quad + (\mathbf{A}_{r,n}^H \mathbf{A}_{r,n}) \odot (\mathbf{B}_{n,n}^{*} \Dot{\mathbf{V}}_{\tau_n,n,n}^H \mathbf{V}_{n,i}).\\
        \left\{\mathbf{F}_{\boldsymbol{\tau}_\textit{n}\mathbf{b}_{m,i}}\right\}_{n\neq m} & = (\mathbf{A}_{r,m}^H \mathbf{A}_{r,m}) \odot (\mathbf{B}_{m,n}^{*} \Dot{\mathbf{V}}_{\tau_n,m,n}^H \mathbf{V}_{m,i}).
    \end{align}
    \label{Eqn::ap_nc7}
\end{subequations}
Lastly, the submatrix component in (\ref{Eqn::ap_nc5(b)}) is written as
\begin{equation}
    \mathbf{F}_{\mathbf{b}\mathbf{b}} = \text{diag}\left(\mathbf{F}_{\mathbf{b}_1\mathbf{b}_1}, \mathbf{F}_{\mathbf{b}_2\mathbf{b}_2}, \ldots, \mathbf{F}_{\mathbf{b}_N\mathbf{b}_N}\right),
\end{equation}
where
\begin{equation}
    \mathbf{F}_{\mathbf{b}_{n,i}\mathbf{b}_{n,j}} = (\mathbf{A}_{r,n}^H \mathbf{A}_{r,n}) \odot (\mathbf{V}_{n,i}^H \mathbf{V}_{n,j}).
\end{equation}
Augmenting all blocks into (\ref{Eqn::ap_nc1}) yields the complete FIM for the noncoherent processing.

\ifCLASSOPTIONcaptionsoff
  \newpage
\fi

% trigger a \newpage just before the given reference
% number - used to balance the columns on the last page
% adjust value as needed - may need to be readjusted if
% the document is modified later
%\IEEEtriggeratref{8}
% The "triggered" command can be changed if desired:
%\IEEEtriggercmd{\enlargethispage{-5in}}

% references section

% can use a bibliography generated by BibTeX as a .bbl file
% BibTeX documentation can be easily obtained at:
% http://mirror.ctan.org/biblio/bibtex/contrib/doc/
% The IEEEtran BibTeX style support page is at:
% http://www.michaelshell.org/tex/ieeetran/bibtex/
\bibliographystyle{IEEEtran}
% argument is your BibTeX string definitions and bibliography database(s)
\bibliography{IEEEabrv,reference}
%
% <OR> manually copy in the resultant .bbl file
% set second argument of \begin to the number of references
% (used to reserve space for the reference number labels box)
%\begin{thebibliography}{1}

%\bibitem{IEEEhowto:kopka}
%H.~Kopka and P.~W. Daly, \emph{A Guide to \LaTeX}, 3rd~ed.\hskip 1em plus
%  0.5em minus 0.4em\relax Harlow, England: Addison-Wesley, 1999.

%\end{thebibliography}

% biography section
% 
% If you have an EPS/PDF photo (graphicx package needed) extra braces are
% needed around the contents of the optional argument to biography to prevent
% the LaTeX parser from getting confused when it sees the complicated
% \includegraphics command within an optional argument. (You could create
% your own custom macro containing the \includegraphics command to make things
% simpler here.)

% if you will not have a photo at all:
%\begin{IEEEbiographynophoto}{John Doe}
%Biography text here.
%\end{IEEEbiographynophoto}

% insert where needed to balance the two columns on the last page with
% biographies
%\newpage

%\begin{IEEEbiographynophoto}{Jane Doe}
%Biography text here.
%\end{IEEEbiographynophoto}

% You can push biographies down or up by placing
% a \vfill before or after them. The appropriate
% use of \vfill depends on what kind of text is
% on the last page and whether or not the columns
% are being equalized.

\vfill

% Can be used to pull up biographies so that the bottom of the last one
% is flush with the other column.
%\enlargethispage{-5in}

% that's all folks
\end{document}